\documentclass[a4paper,fleqn,usenatbib,useAMS]{mnras}

\usepackage{graphicx}	
\usepackage{amsmath}	
\usepackage{amssymb}	
\usepackage{multicol}        
\usepackage{bm}		
\usepackage[T1]{fontenc}
\usepackage{ae,aecompl}
\usepackage{newtxtext,newtxmath}
\usepackage[table,dvipsnames]{xcolor}			
\usepackage[normalem]{ulem}

\title[Shear-driven magnetic buoyancy]{Shear-driven magnetic buoyancy in the solar tachocline: The mean electromotive force due to rotation}
\author[C. D. Duguid et al.]{
	Craig. D. Duguid,\thanks{E-mail: craig.duguid@newcastle.ac.uk}
	Paul J. Bushby,
	and Toby S. Wood 
	\\
	School of Mathematics, Statistics and Physics, Newcastle University, Newcastle Upon Tyne, NE1 7RU, UK\\
}

\date{Accepted XXX. Received YYY; in original form ZZZ}
\pubyear{2022}

\begin{document}
\label{firstpage}
\pagerange{\pageref{firstpage}--\pageref{lastpage}}
\maketitle

\begin{abstract} 
The leading theoretical paradigm for the Sun’s magnetic cycle is an $\alpha\omega$-dynamo process, in which a combination of differential rotation and turbulent, helical flows produces a large-scale magnetic field that reverses every 11 years.  
Most $\alpha\omega$ solar dynamo models rely on differential rotation in the solar tachocline to generate a strong toroidal field. 
The most problematic part of such models is then the production of the large-scale poloidal field, via a process known as the $\alpha$-effect. 
Whilst this is usually attributed to small-scale convective motions under the influence of rotation, the efficiency of this regenerative process has been called into question by some numerical simulations.
Motivated by likely conditions within the tachocline, the aim of this paper is to investigate an alternative mechanism for the poloidal field regeneration, namely the magnetic buoyancy instability in a shear-generated, rotating magnetic layer.
We use a local, fully compressible model in which an imposed vertical shear winds up an initially vertical magnetic field.  
The field ultimately becomes buoyantly unstable, and we measure the resulting mean electromotive force (EMF).  
For sufficiently rapid rotation, we find that a significant component of the mean EMF is aligned with the direction of the mean magnetic field, which is the characteristic feature of the classical $\alpha\omega$-dynamo model.  
Our results therefore suggest that magnetic buoyancy could contribute directly to the generation of large-scale poloidal field in the Sun.
\end{abstract}

\begin{keywords}
MHD -- Sun: magnetic fields -- dynamo -- Sun: rotation -- hydrodynamics -- instabilities
\end{keywords}

\section{Introduction}\label{section_introduction}
The key properties of the solar magnetic cycle, which waxes and wanes with a period of approximately 11 years, are well known. However, the nature of the dynamo mechanism that is responsible for producing this cyclic behaviour is still not fully understood \citep{Charbonneau_living_review}.
One of the most plausible explanations for the observed magnetic activity is that of an $\alpha\omega$-dynamo \citep[first proposed by][]{parker_hydromagnetic_1955}.
A key component of this dynamo mechanism is differential rotation (the $\omega$-effect), which stretches magnetic field lines in the direction of the flow; around the base of the solar convection zone, the strong radial shear in the solar tachocline \citep{thompson_internal_2003} would tend to produce a magnetic field with a dominant azimuthal (toroidal) component.
For a successful dynamo, there must also be some mechanism that regenerates the poloidal (radial and latitudinal) components of the large-scale magnetic field.
In Parker's scenario \citep{parker_hydromagnetic_1955,parker_solar_1993}, the dynamo loop is completed by the action of cyclonic convection upon toroidal field lines (a process now referred to as the $\alpha$-effect).
Due to the action of the Coriolis force, convective upwellings tend to produce rising, twisted magnetic loops.
The cumulative effect of many such loops could, in principle,
regenerate a large-scale poloidal field. 
Using a model in which the $\alpha$-effect was implemented via a parametrised source term, Parker established that an $\alpha\omega$-dynamo of this type should be capable of producing oscillatory magnetic behaviour, similar to that seen in the Sun, provided that the $\alpha$- and $\omega$-effects are efficient enough to overcome ohmic dissipation.  

While there is a strong consensus, driven by the observations, that the $\omega$-effect is at work in the solar interior, the $\alpha$-effect remains a controversial area.
It is possible to place Parker's heuristic arguments on a firmer mathematical footing using the techniques of mean-field electrodynamics \citep{steenbeck-etal66,moffatt_magnetic_1978}.
Under this approach, an averaging process is introduced, and the equation for the mean magnetic field then contains an additional source term (a mean electromotive force, EMF) that arises as a result of the correlations between the fluctuating parts of the magnetic field and the flow.
For systems that lack reflectional symmetry (e.g. rotating systems), Parker's $\alpha$-effect then emerges (under certain assumptions) as part of this mean EMF.
By making the first-order smoothing approximation \citep[see, e.g.,][for more details]{moffatt_magnetic_1978}, it is possible to derive an expression for the $\alpha$-effect that is directly proportional to the kinetic helicity of the flow. 
However, the validity of this expression is questionable under the highly-conducting, turbulent conditions
present in 
the solar interior. 

Even in (non-solar-like) situations where there is no differential rotation, mean-field theory suggests that rapidly rotating convection should be capable of sustaining a magnetic field with a significant large-scale component (i.e. a magnetic field that is structured on the scale of the system).
There are certainly examples of large-scale ($\alpha^2$) dynamos in near-onset, rapidly-rotating convection \citep[e.g.][]{childress_convection-driven_1972} and in moderately supercritical, rotationally-dominated convection \citep[e.g.][]{masada_spontaneous_2016,bushby_large_scale_2018}.
However, dynamos driven by turbulent rotating convection tend to produce disordered, small-scale magnetic fields \citep[e.g.][]{cattaneo_dynamo_2006,favier_bushby_2013}.
These results are often interpreted in terms of the $\alpha$-effect, and it is certainly possible to measure $\alpha$ in numerical simulations of convection in a rotating domain \citep[see, e.g.][]{cattaneo_dynamo_2006,kapyla_large_scale_2009,favier_bushby_2013}.
However, the outcome of such measurements is dependent upon the techniques used, and this remains an area of some disagreement in the literature (a full discussion of which lies beyond the scope of the present paper).
Nonetheless, these studies cast doubt on whether 
cyclonic convection can produce the required $\alpha$-effect in the turbulent conditions of the solar convection zone.

Given the above considerations, it is natural to ask whether there are any other physical mechanisms that could give rise to a similar ``rise and twist'' effect that Parker originally envisaged as being due to cyclonic convection (thus completing the dynamo cycle).
One candidate for this is magnetic buoyancy. It is well established that isolated magnetic flux tubes tend to be less dense than their non-magnetic surroundings, and are therefore buoyant \citep{parker_formation_1955,jensen_tubes_1955}, and this has long been cited as an explanation for the emergence of
magnetically
active regions at the solar surface.
Whilst it has been shown that instabilities in individual flux tubes can (in the presence of rotation) give rise to a mean EMF that is analogous to the $\alpha$-effect \cite[][]{ferriz_mas_dynamo_1994}, it is unlikely that the magnetic field distribution in the solar interior takes the form of discrete, isolated magnetic flux tubes.
It is more plausible that there is a continuous layer of predominantly azimuthal magnetic flux that is \citep[partly as a consequence of flux pumping, e.g.][]{tobias_transport_2001} largely confined to the sub-adiabatically stratified region just below the base of the solar convection zone, where poloidal field can be stretched out by the shear in the tachocline. 

The evolution of an imposed magnetic layer under the action of magnetic buoyancy has been well studied \citep[e.g.][]{gilman_instability_1970,cattaneo_nonlinear_1988,matthews_magnetic_1995,wissink_three-dimensional_2000,kersale_nonlinear_2007,mizerski_short_wavelength_2013,gilman_magnetic_2018}.
The preferred mode of instability is usually three-dimensional, producing undular motions with a long wavelength parallel to the magnetic field lines and a short wavelength perpendicular to them.
As noted by \citet{hughes_solar_2007}, for example, a simple comparison of the typical growth rate of the instability compared to the rotation period of the Sun suggests that rotation should be playing a role in the evolution of this instability at the base of the solar convection zone.
The effects of rotation upon the magnetic buoyancy instability are non-trivial \citep[see, e.g.][for a review]{hughes_solar_2007}.
Here we focus upon those studies of magnetic buoyancy in the presence of rotation that are directly relevant to the
solar
dynamo problem. 
Building on the analysis of \citet{gilman_instability_1970}, \citet{moffatt_magnetic_1978} considered the magnetic buoyancy instability in a unidirectional layer under the magnetostrophic approximation; he was able to show that the resultant instability produces a systematic mean EMF from which it is possible to derive an $\alpha$-effect.
The mean electromotive force due to magnetic buoyancy in the presence of rotation (and its possible influence on the solar dynamo) has been explored in a number of subsequent studies \citep{Schmitt_1984,brandenburg_simulations_1998,thelen_mean_2000,thelen_nonlinear_2000,chatterjee_alpha_2011,davies_mean_2011}. 
It should be noted that \citet{davies_mean_2011} suggest that attention should be focused upon the mean EMF in such cases, rather than the $\alpha$-effect itself, arguing that the decomposition of this quantity into standard mean-field coefficients is too simplistic to be meaningful in the context of this magnetically-driven instability.  

All of the magnetic buoyancy studies listed above consider the evolution of an imposed magnetic layer under the action of this instability. 
Inevitably, the evolution of this layer will depend to some extent on the initial magnetic field distribution.
In the solar tachocline, it is believed that this layer is generated by shearing motions, so it is clearly of importance (as a step towards the full dynamo problem) to consider the corresponding magnetic buoyancy instability in a shear-generated magnetic layer.
In a series of papers, \citet{brummell_formation_2002}, \citet{cline_formation_2003} and \citet{cline_dynamo_2003} considered the evolution of buoyant magnetic structures, generated by an idealised shear flow with a strong horizontal (latitudinal) gradient.
Although the shear flow was not particularly tachocline-like, they did manage to generate dynamo action \citep{cline_dynamo_2003}, even in the absence of rotation, via the combination of magnetic buoyancy and Kelvin-Helmholtz instabilities.
The more solar-like problem of a magnetic layer generated by a vertical shear has also been considered \citep[][]{vasil_magnetic_2008,vasil_constraints_2009,silvers_interactions_2009,silvers_double-diffusive_2009,barker_magnetic_2012}. In their initial study, \citet{vasil_magnetic_2008} found that it was only possible to excite a magnetic buoyancy instability in this configuration with an unrealistically strong (hydrodynamically unstable) shear.
A resolution to this problem was later proposed by \citet{silvers_double-diffusive_2009}, who observed that it was in fact possible for such systems to produce magnetic buoyancy instabilities, even with a comparatively weak shear, if the ratio of the magnetic to thermal diffusivities is sufficiently small.
Fortunately, this is exactly the expected parameter regime for the solar tachocline.

In terms of modelling the solar dynamo, the key ingredient that is missing from previous studies that have considered magnetic buoyancy in a shear-generated layer is rotation.
We know from the imposed field calculations that magnetic buoyancy instabilities can produce a mean EMF that should be conducive to dynamo action.
What we do not yet know is whether or not a suitable mean
EMF can be obtained from the shear-generated magnetic buoyancy instability.
This is the subject of the present paper and this represents a crucial step towards building a magnetic buoyancy-driven version of Parker's solar dynamo model. 

The plan for this paper is as follows. In Section 2 we will describe the model and governing equations.
In Section 3 we will report on the results of the hydrodynamic problem on the stability of the shear for non-rotating and rotating systems. 
In section 4 we analyse results of simulations with and without rotation for the full MHD problem. 
We will then present the most important results of this work in Section 5, namely, the mean EMF driven by this magnetic buoyancy instability.
We will then discuss the importance of our results in Section 6, focusing particularly upon the potential for this system to act as a dynamo.

\section{Model}\label{section_model}
\subsection{Model set-up}

\begin{figure}
	\begin{center}
    \includegraphics[width=0.8\columnwidth]{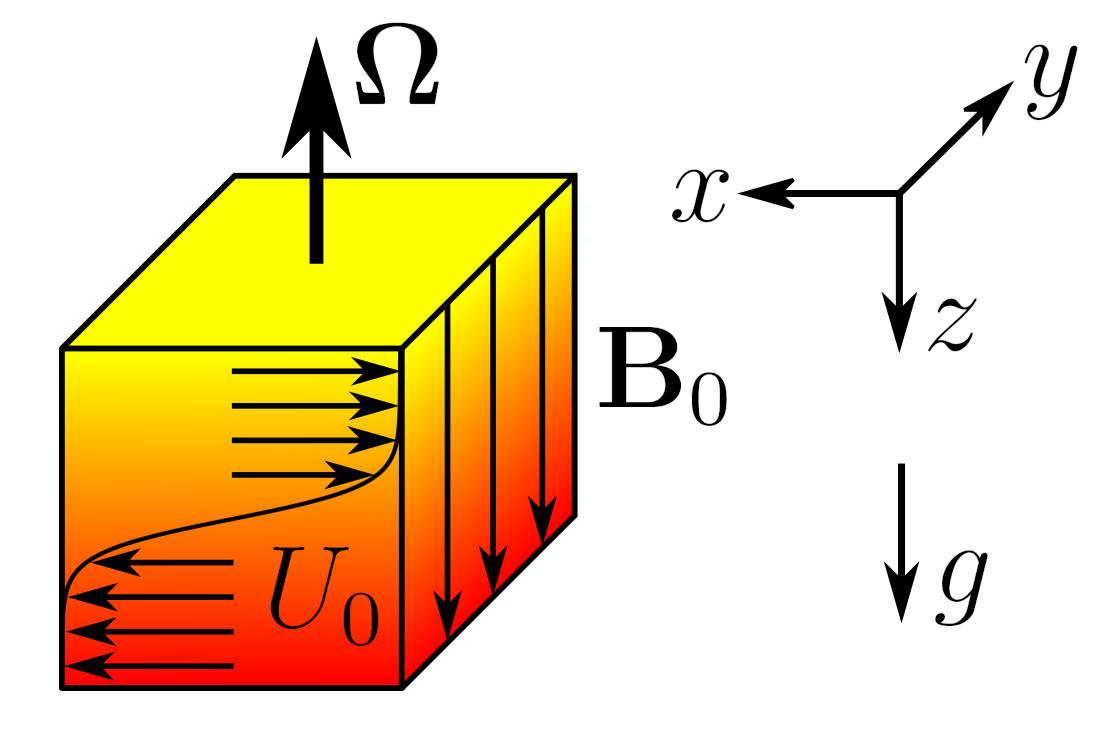}
	\caption{Schematic of the model domain. The arrows in the $x$-direction show the shear flow, $U_0$, whilst the arrows in the $z$-direction show the initial magnetic field, $\mathbf{B}_0$.
	The shading indicates the initial temperature distribution, with red (yellow) shading corresponding to warmer (cooler) fluid.}
	\label{fig_geometry_and_shear}
	\end{center}
\end{figure}

A schematic of the model geometry is shown in Fig.~\ref{fig_geometry_and_shear}.
We consider a Cartesian domain, of depth $d$, with the coordinate system oriented so that the $z$-axis points vertically downwards, parallel to the constant gravitational acceleration, $g\mathbf{e}_z$. 
This domain rotates uniformly about the $z$-axis, with a constant angular velocity, $-\Omega\mathbf{e}_z$.
The domain 
is filled with a compressible, electrically-conducting fluid, of constant thermal conductivity $K$, constant magnetic diffusivity $\eta$, and constant dynamic viscosity $\mu$.
The (constant) magnetic permeability of this fluid is denoted by $\mu_0$.
This fluid is assumed to be an ideal gas, with constant specific heats $c_p$ and $c_v$ ($\mathfrak{R}=c_p-c_v$ is the gas constant). 
In its initial state, this fluid is horizontally-uniform, with temperature, $T_0$, and density, $\rho_0$, at the upper surface ($z=0$). 
A fixed heat flux is applied through the lower boundary ($z=d$), leading to an initial temperature difference across the domain of $\Delta T$. 
The initial state is that of a polytrope, with 
\begin{align}
	T(z) = T_0\left(1 + \frac{\theta z}{d}\right) \,, \quad \text{and} \quad \rho(z) = \rho_0\left(1 + \frac{\theta z}{d}\right)^m\,,
\end{align}
where $\theta = \Delta T/T_0$ is a measure of the thermal stratification of the domain, and $m=(gd/\mathfrak{R}\Delta T)-1$ is the polytropic index.

Following previous studies \citep{vasil_magnetic_2008,silvers_double-diffusive_2009}, we include an additional forcing term in the momentum equation.
The aim is to mimic the key aspects of the differential rotation in a local region of the solar tachocline.
Identifying the $x$-axis of the domain with the local azimuthal (toroidal) direction, this shear flow takes the form $\mathbf{u} = U_0(z)\mathbf{e}_x$, where $U_0(z)$ will be specified below. 
The forcing is chosen so as to balance this flow (in the absence of a magnetic field), and this flow is set as an initial condition across the polytropic layer. 

Having set up the polytropic state with a shear flow, we then introduce a seed magnetic field into the system. 
This field is initially uniform and vertical, taking the form $\mathbf{B}=B_0\mathbf{e}_z$.
This imposed field will be continually stretched in the $x$-direction by the shear, producing a horizontal magnetic layer (the peak strength of which initially increases linearly with time).
Eventually, the field in this layer will be amplified to a level at which it becomes dynamically significant, resisting the shearing motions and ultimately driving a magnetic buoyancy instability in the system. 

Given the expected geometry of the magnetic layer, we anticipate that the ensuing magnetic buoyancy instability will have a long length-scale in the $x$-direction and a short length-scale in the $y$ direction.
Following \citet{silvers_double-diffusive_2009}, we choose to reflect the anisotropic nature of the instability in the domain geometry. 
Defining the horizontal dimensions to be $0 \le x \le L_x d $ and $0\ \le y \le L_y d$, we set $L_x=2$ and $L_y=0.5$. 
This narrow domain enables us to study the key features of the system without expending unnecessary computational effort. 

It is worth emphasising at this stage that it is the inclusion of rotation that represents the novel aspect of the present study. 
To the best of our knowledge, this is the first time that the effects of rotation have been included in a study of magnetic buoyancy in a shear-generated layer. 
As we shall see later, this raises some interesting issues in terms of the shear forcing, and also leads to the excitation of mean flows once the magnetic field becomes dynamically significant. 
Given the additional complexity that this introduces into the system, we chose to focus exclusively upon the simplest case of a vertical rotation vector in this initial study (which places this domain within the polar regions of the solar tachocline). 
We will explore the case of an inclined rotation vector, allowing us to probe the latitudinal dependence of the system, in a future paper.

\subsection{Governing equations}

In order to make the governing equations dimensionless, we adopt similar scalings to those set out in several previous studies \citep[e.g.][]{matthews_compressible_1995,favier_bushby_2013}.
The characteristic length-scale is assumed to be the depth of the layer, $d$, whilst the density and temperature are scaled in terms of their initial values at the upper surface of the domain ($\rho_0$ and $T_0$ respectively).
Our adopted time-scale is the (isothermal) acoustic travel time at the top of the domain ($z=0$) defined by $d/\sqrt{\mathfrak{R}T_0}$.
A natural scaling for the velocity of the fluid is then the corresponding isothermal sound speed, $\sqrt{\mathfrak{R}T_0}$, whilst the magnetic field is scaled in terms of $B_0$.

In these units, the equations describing the temporal evolution of the density $\rho$, temperature $T$, velocity $\mathbf{u}$ and magnetic field $\mathbf{B}$ become

\begin{subequations}\label{maths_p3d_equations}
	\begin{align}
		\rho \dfrac{\partial  \mathbf{u}}{\partial t} + \rho (\mathbf{u} \cdot \nabla) \mathbf{u}&= -{\mathrm{Ta}_0}^{1/2} \, \sigma \kappa \rho \boldsymbol{\Omega}\times \mathbf{u} - \nabla P + \theta(m+1) \rho \mathbf{e}_z   \nonumber \\
		& \quad \quad + F(\nabla \times \mathbf{B})\times \mathbf{B} + \sigma \kappa \nabla \cdot \tau + \mathbf{F}_s \,,\label{maths_momentum}\\
		\rho\dfrac{\partial T}{\partial t} + \rho (\mathbf{u} \cdot \nabla) T & = -(\gamma -1) P \nabla \cdot \mathbf{u} + \gamma \kappa \nabla^2 T \nonumber \\
		& \quad \quad + F ( \gamma -1) \zeta_0 \kappa \lvert\nabla\times\mathbf{B} \rvert^2 + \frac{(\gamma-1) \sigma \kappa}{2} \tau^2   \,,\label{maths_temperature_equation}\\
		\dfrac{\partial \mathbf{B}}{\partial t} & = \nabla \times (\mathbf{u} \times \mathbf{B} - \zeta_0 \kappa \nabla \times \mathbf{B}) \,,\label{maths_induction}\\
		\dfrac{\partial \rho}{\partial t} &= - \nabla \cdot (\rho \mathbf{u}) \,,\label{maths_continuity}\\ 
		\nabla \cdot \mathbf{B} & = 0 \label{maths_solenoidal_magnetic_field}\,.  
	\end{align}
\end{subequations}
We have defined the viscous stress tensor as
\begin{align}
	\tau_{ij} & = \dfrac{\partial u_i}{\partial x_j} + \dfrac{\partial u_j}{\partial x_i} - \frac{2}{3} \delta_{ij} \dfrac{\partial u_k}{\partial x_k} \label{maths_stress_tensor}
\end{align}
(where $\delta_{ij}$ denotes the Kronecker delta), whilst the pressure $P$ satisfies the equation of state for an ideal gas, 
\begin{align}
  P=\rho T\,.
\end{align}
This system contains a number of non-dimensional parameters,
which are summarised in Table~\ref{table_dimensionless_parameters} along with their typical values. These are discussed in more detail below.

In Eq.~(\ref{maths_momentum}), the forcing term, $\mathbf{F}_s$, is chosen to balance the viscous and Coriolis forces associated with the imposed shear flow, $U_0(z)\mathbf{e}_x$.
In the absence of magnetic field or any hydrodynamic instabilities, we would therefore expect the fluid to remain in a steady state with $\mathbf{u} = U_0(z)\mathbf{e}_x$.
Following a similar approach to \citet{vasil_magnetic_2008} and \cite{silvers_double-diffusive_2009} we choose a shear flow of the form
\begin{equation}
    U_0(z) = A \tanh [ 10(z-0.5) ]  \,,
\end{equation}
where the constant $A$ determines the shear amplitude.
The form of the shear flow is illustrated schematically in Fig.~\ref{fig_geometry_and_shear}. 
This hyperbolic tangent profile localises the shearing region at the mid-plane of the domain. 
We note that, for sufficiently large values of $A$, the shear flow would be unstable even in the absence of a magnetic field; the hydrodynamic stability of this system (and the consequent choice of $A$) will be discussed in \S~\ref{section_shear_stability}.
In order to maintain this shear flow, we take
\begin{align} \label{maths_shear_forcing}
\mathbf{F}_s & =  \begin{bmatrix}
           - \sigma \kappa \partial_{zz} U_0 \\
           \sqrt{\mathrm{Ta}_0}\, \sigma \kappa \rho U_0  \\
           0
         \end{bmatrix} \,. 
\end{align}
In the $x$-direction, this forcing balances the viscous term. 
Note that the presence of rotation requires $\mathbf{F}_s$ to have a $y$ component in order to balance the Coriolis contribution from the imposed shear.
Once the magnetic field is introduced into the fluid, the Lorentz tension in the field lines will eventually act to inhibit the shear, and may also drive other kinds of mean flow.
However, by making the initial magnetic field sufficiently weak, it is possible minimise the influence of some of these dynamical effects upon the evolution of the system over the timescales of interest.  

\subsection{Boundary and initial conditions}

All variables are assumed to be periodic in both of the horizontal directions. 
The upper and lower bounding surfaces are assumed to be stress-free and impermeable. 
The upper boundary is held at fixed temperature, whilst (as already noted) a fixed heat flux is imposed through the lower boundary. 
On both bounding surfaces, the horizontal magnetic field is assumed to vanish.
Therefore the tangential viscous, Reynolds
and Maxwell stresses all vanish on the upper and lower boundaries.
In this dimensionless system, these conditions correspond to 

\begin{subequations}\label{maths_bounday_conditions}
\begin{gather}
	u_z = \dfrac{\partial u_x}{\partial z} = \dfrac{\partial u_y}{\partial z} = 0  \quad \text{for} \quad z\in\{0,1\}\,,\\
	T(z=0) = 1   \quad \text{and} \quad  \dfrac{\partial T}{\partial z}\Big\lvert_{z=1} = \theta \,,\\
	B_x = B_y = \dfrac{\partial B_z}{\partial z} = 0 \quad \text{for} \quad z\in\{0,1\}\,.
\end{gather}
\end{subequations}
Note that the shear is sufficiently localised about the mid-plane that there is no contradiction here in terms of the stress-free condition on the velocity field. 
It is straightforward to confirm that these boundary conditions are consistent with the initial polytropic solution, which (in the absence of a magnetic field) is then an equilibrium solution of the governing equations.
This equilibrium is eventually perturbed by the magnetic field as it is amplified by the shear. 
However, a further perturbation is needed in order to seed the magnetic buoyancy instability. 
Therefore, a small thermal perturbation is also added to the initial state for each simulation. 
This takes the form of a small, pseudo-random perturbation to the temperature distribution (additive noise, localised around the mid-plane, of peak amplitude $0.05$ in dimensionless units).

\begin{table}
	\begin{center}
\def\arraystretch{2}
	\begin{tabular}{ c | c | c | c}
		Symbol & Description & Definition  & Values \\ \hline
		$F$ & Magnetic field strength & $\dfrac{B_0^2}{\mathfrak{R} T_0 \rho_0 \mu_0}$ & Variable \\
		$\sigma$ & Prandtl number & $\dfrac{\mu c_p}{K}$ & 0.00025  \\
		$\theta$ & Temperature gradient & $\dfrac{\Delta T}{T_0}$ & 5 \\
		$\kappa$ & Thermal diffusivity & $\dfrac{K}{d \rho_0 c_p \sqrt{\mathfrak{R}T_0}}$ & 0.01	 \\
		$\zeta_0$ & Inverse Roberts number & $\dfrac{\eta c_p \rho_0}{K}$  & 0.0005 \\
		$\gamma$ & Ratio of specific heats & $\dfrac{c_p}{c_v}$ & $5/3$ \\
		$m$ & Polytropic index & $\dfrac{g d}{\mathfrak{R} \Delta T} -1$  & 1.6 \\
		$\mathrm{Ta}_0$ & Taylor number &  $\dfrac{4\rho_0^2 \Omega^2 d^4}{\mu^2}$  & Variable  \\  \hline
		$\mathrm{Re}$ & Reynolds number & $\dfrac{U_0 \rho_0}{\sigma \kappa}$ & 8000  \\
		$\mathrm{Rm}$ & Magnetic Reynolds number & $\dfrac{U_0}{\zeta_0 \kappa}$ & 4000 \\
		$\sigma_m$ & Magnetic Prandtl number & $\dfrac{\sigma}{\zeta_0}$ & 0.5  \\		\hline
		$\tau_\text{visc}$ & Viscous timescale & $\dfrac{1}{\sigma \kappa}$ & $4\times10^{5}$ \\
		$\tau_\text{ohmic}$ & Ohmic timescale & $\dfrac{1}{\zeta_0 \kappa}$ & $2\times10^{5}$ \\
		$\tau_\text{therm}$ & Thermal timescale & $\dfrac{1}{\kappa}$ & $100$ \\
	\end{tabular}
\caption{Non-dimensional parameters in the system including a text description/name of the quantity, the definition, and the value the parameter takes (where applicable). 
Also reported are the (shear) fluid and magnetic Reynolds numbers (assuming a shear amplitude of $A=0.02$), the magnetic Prandtl number, and the diffusive timescales (viscous, Ohmic, and thermal). These Reynolds numbers and timescales are defined using the total layer depth as the characteristic length-scale (which is approximately one order of magnitude larger than the initial width of the shear layer).} 
\label{table_dimensionless_parameters}
\end{center}
\end{table}	

\subsection{Non-dimensional parameters}\label{section_parameters} 
In Table~\ref{table_dimensionless_parameters} we list the non dimensional parameters with their typical values \citep[which largely follows][]{silvers_double-diffusive_2009}. 
Most of the parameters are fixed. 
In particular, we choose the Prandtl number, the thermal diffusivity and the inverse Roberts number so that the viscous, Ohmic and thermal diffusion timescales have the correct ordering for the tachocline, with the thermal timescale much shorter than the other two, and the magnetic Prandtl number less than unity
(although because of computational limitations these simulations are necessarily much more dissipative than the real tachocline). 
The comparatively short thermal diffusion time tends to promote magnetic buoyancy in this system
\citep{silvers_double-diffusive_2009}. 
With $\gamma=5/3$, a polytropic index of $m=1.6$ ensures that the layer is sub-adiabatically stratified. With this choice of parameters, our domain covers $N_p = (m+1) \ln(1+\theta) \approx 4.66$ pressure scale heights. 
In terms of the parameters to be varied, the quantity $F$ gives the ratio of the squared Alfv\'en speed to the squared (isothermal) sound speed at $z=0$ at the start of the calculation. 
Note that a multiplicative factor of $\sqrt{F}$ must be included in front of the magnetic field in order to make quantitative comparisons between the magnetic field strength and the flow. 
With all other parameters fixed, varying $F$ is equivalent to varying the strength of the initial vertical magnetic field. The other parameter to be varied is the Taylor number, $\mathrm{Ta}_0$, which is a dimensionless measure of the rotation rate of the system relative to viscous dissipation. 
Due to the dependence of the Coriolis term upon $\rho$, it is arguably the mid-layer value of this ratio that is of direct relevance to the shear (and hence the magnetic layer); we therefore quote the mid-layer Taylor number in what follows, i.e.

\begin{equation}
    \mathrm{Ta} = \mathrm{Ta}_0\left(1+\frac{\theta}{2}\right)^{2m} =\dfrac{4\rho_0^2 \Omega^2 d^4}{\mu^2}\left(1+\frac{\theta}{2}\right)^{2m} \,.
\end{equation}
With this choice of parameters, $\mathrm{Ta}\approx 55\mathrm{Ta}_0$.

\subsection{Numerical method}

We solve the system of governing equations (Eq.~\ref{maths_p3d_equations}) numerically. 
In practice, we use a poloidal-toroidal decomposition for the magnetic field, which ensures that it remains solenoidal at all times. 
Horizontal derivatives are computed in Fourier space, whereas a 4th-order finite difference scheme is applied to the vertical derivatives (in order to improve stability, an upwind scheme is used for the advective terms, where appropriate). 
All variables are de-aliased in Fourier space using the standard 2/3 rule. 
In order to time-step the equations, we use an explicit third-order Adams-Bashforth scheme with a variable time-step. 
We use a resolution of $(N_x, N_y, N_z) = (192, 96, 192)$ in all cases presented here. 
We have also performed a more extensive low resolution parameter survey with a resolution of $(N_x, N_y, N_z) = (128, 64, 128)$, the results of which are not presented in the main body of the paper. 
However, it should be noted that the differences between the lower and higher resolutions are not significant. 
A full list of all simulations performed in this work can be found in Appendix \ref{appendixA}.

\section{Hydrodynamic considerations}\label{section_shear_stability}
We first consider the stability of the shear flow in the absence of a magnetic field. 
The purpose of this is to ensure that the dynamics that we observe in the subsequent magnetic calculations are entirely driven by magnetic buoyancy, as opposed to an underlying hydrodynamic instability. 

Focusing initially upon the hydrodynamic system in the absence of rotation, we note that the Richardson number is defined as follows:

\begin{align}
	\mathrm{Ri} &= \frac{N^2}{(\mathrm{d}U_0/\mathrm{d}z)^2} \nonumber \\
	&= \frac{\theta^2(m+1)}{100A^2} \left( m -\frac{1}{\gamma}\left[m+1 \right] \right) \frac{\cosh^4\Big(10 (z - 0.5) \Big)}{1 + \theta z}\,,
\end{align}
where $N$ is the Brunt-V\"{a}is\"{a*}l\"{a} frequency, which measures the strength of stratification, and $\mathrm{d}U_0/\mathrm{d}z$ is the local shearing rate. 
At fixed $\theta$, $m$ and $\gamma$, the key point to note is that this quantity is inversely proportional to the square of the shear flow amplitude, $A$.
In the absence of diffusion, a necessary condition for stability is that $\mathrm{Ri}>1/4$ is satisfied everywhere in the domain \citep{miles_stability_1961,howard_note_1961}. 
This condition was not satisfied in the strong shear regime considered by \citet{vasil_magnetic_2008}, who found a vigorous hydrodynamic instability in the absence of a magnetic field. 
However, this condition was satisfied in the calculations of \cite{silvers_double-diffusive_2009}, who quote a value of $\mathrm{Ri}\approx 2.96$. 
When diffusive effects are included, some care is needed in the limit of rapid thermal diffusion, which can have a destabilising influence upon stratified shear flows \citep{ledoux_rotational_1974,garaud_turbulent_2017,cope_dynamics_2021}. 
In this case, the stability of the system depends not only upon the Richardson number, but also the Prandtl number. 
Building on the work of \cite{ledoux_rotational_1974}, \citet{garaud_turbulent_2017} find a condition for instability that $\mathrm{Ri}\,\sigma \lesssim 0.007$. 
Whilst this threshold is empirically determined, probably exhibiting some degree of dependence upon the details of the model, it is notable that the system considered by \citet{silvers_double-diffusive_2009} yields a value of $\mathrm{Ri}\,\sigma \approx 0.00074$, which is well below the stability threshold suggested by \citet{garaud_turbulent_2017}. 
In other words, their underlying shear is potentially (hydrodynamically) unstable in this parameter regime.  

In order to investigate this issue, we consider here two shear amplitudes (initially in the absence of rotation). 
Following \citet{silvers_double-diffusive_2009}, the stronger shear corresponds to $A=0.05$. 
We also consider a weaker shear of $A=0.02$. 
The choice of $A=0.02$ is something of a compromise. 
Whilst $\mathrm{Ri}\approx 18.5 \gg 1/4$ for the case of $A= 0.02$, $\mathrm{Ri}\,\sigma \approx 0.004$, which is still just below the empirically-determined stability threshold of \citet{garaud_turbulent_2017}. 
However, the weaker the shear, the longer the time taken for the formation of the horizontal magnetic layer (during the early stages of evolution, the peak value of $B_x$ is proportional to $At$), so longer (computationally expensive) integrations are needed before the layer becomes buoyantly unstable. 

\begin{figure}
	\includegraphics[width=\columnwidth]{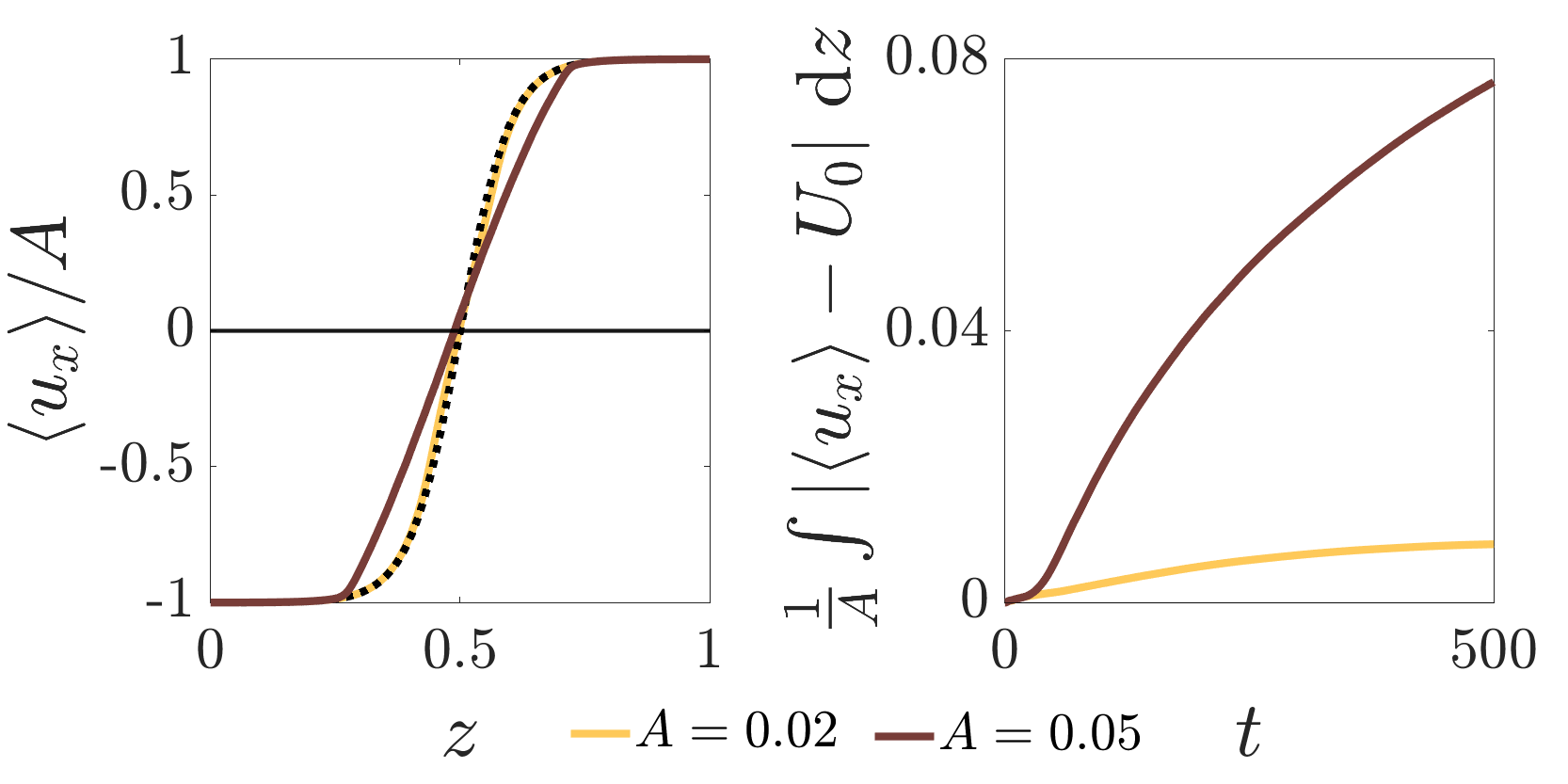}
	\caption{Two metrics for the estimated stability of the shear for the shear parameters of \protect\cite{silvers_double-diffusive_2009} ($A=0.05$) and the shear considered in this work ($A=0.02$), see legend for colours. Left: the normalised (horizontally-averaged) vertical shear profile taken at $t=500$ time units in each case. The dotted line represents the normalised target shear profile. Right: the volume averaged deviation between the normalised shear flow and the normalised target profile.}
	\label{fig_hydrostability}
\end{figure}

The effects of varying the shear amplitude are illustrated in Fig.~\ref{fig_hydrostability}. 
This figure shows the evolution of the horizontally-averaged $x$-component of the velocity field, 

\begin{align}
	\langle u_x \rangle(z,t) = \frac{1}{L_x L_y} \int_0^{L_y} \int_0^{L_y} u_x \, \text{d}x\, \text{d}y
\end{align}
(angled brackets will be used throughout this paper to denote a horizontal average of this form).
In particular, this figure shows the deviation from the initial shear profile for the two shear amplitudes considered, alongside the integrated deviation from the initial profile.  
If even a weak instability is present then one would expect the shear profile to deviate from the target profile as energy is extracted from the shear. 
After $500$ time units (which will be the typical period of time over which we evolve the magnetic buoyancy calculations), we see that the stronger shear ($A=0.05$) has departed significantly from the target shear and the deviations are still increasing with time. 
This suggests that this shear flow is in fact (weakly) unstable. 
On the other hand, the weaker shear profile still shows excellent agreement with the target shear, with an integrated error that appears to be reaching a steady state of approximately $1\%$. 
Over the timescales of interest the shear is essentially unchanged, so we choose $A=0.02$ throughout in what follows; this choice should ensure that any departures from the target shear are definitely magnetically-driven. 
It is worth noting here that the vertical magnetic field we subsequently introduce will
(at least initially) tend to inhibit any shear-driven instabilities \citep[e.g.][]{silvers_interactions_2009}, further reducing the possible impact of any underlying hydrodynamical instability.

\begin{figure}
	\includegraphics[width=\columnwidth]{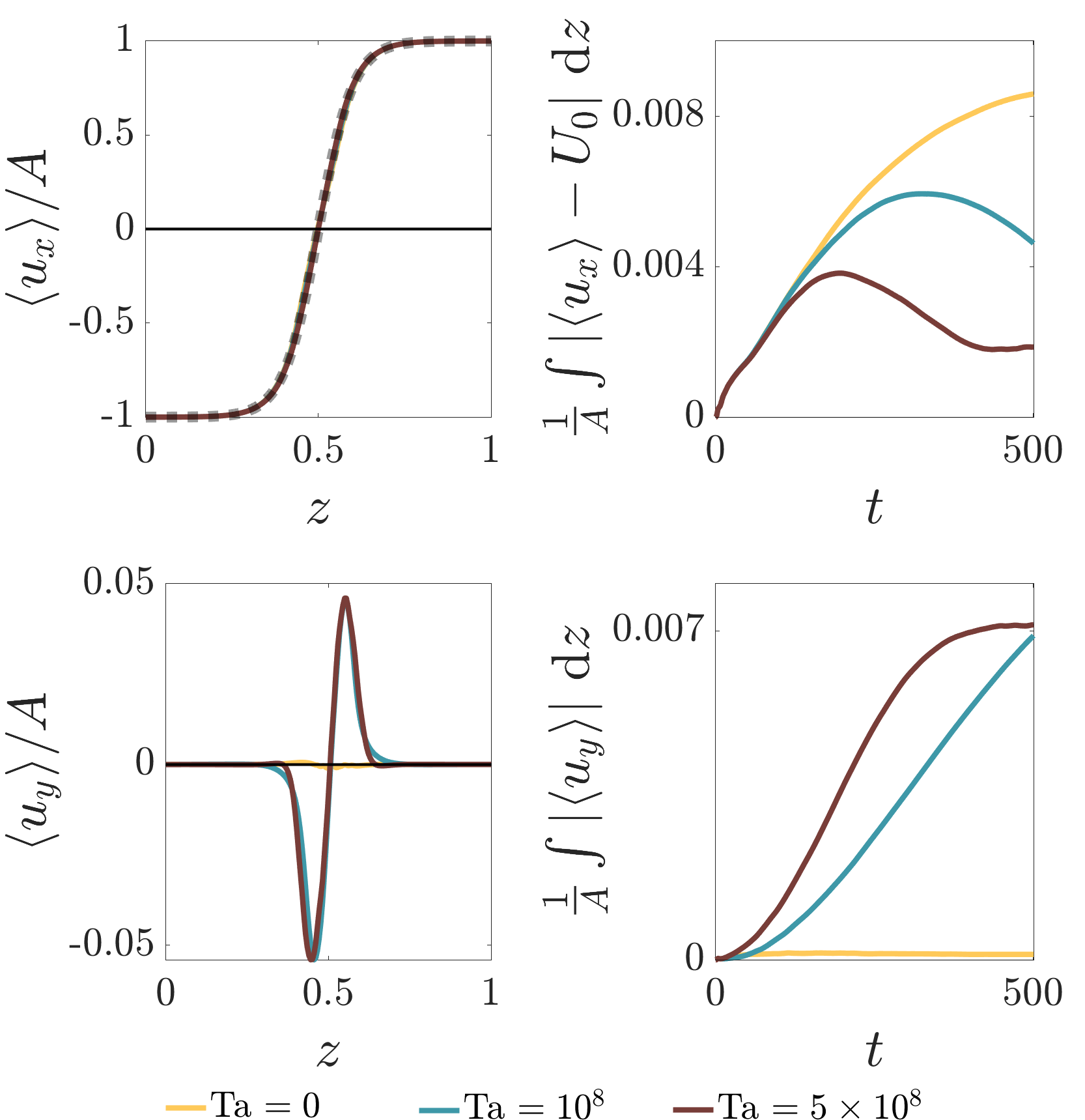}
	\caption{The evolution of the mean flows (for $A=0.02$) for three different rotation rates, with $\text{Ta} \in \{0,1,5\}(\times 10^{8})$ as denoted by the legend. Top left: snapshots of the horizontally averaged vertical shear profile normalised by the shear amplitude for the three rotation rates taken at $t=500$.
	The initial shear profile, $U_0(z)$, is overplotted as a dotted line, and is practically indistinguishable.
	Top right: the volume averaged deviation between the normalised shear flow and the normalised target profile. Bottom left: snapshots of the $\langle u_y \rangle$ profile for each rotation rate taken at $t=500$. Bottom right: the volume integrated deviation of $\langle u_y\rangle$ from zero.}
	\label{fig_hydrostability_rotating}
\end{figure}

The discussion of shear stability has so far been in the context of non-rotating systems, and we need to confirm that our conclusions are unchanged if the system is rotating. 
It is also important to check that our imposed forcing correctly balances the initial shear when Coriolis effects are present.
Fixing the shear amplitude to $A=0.02$, Fig.~\ref{fig_hydrostability_rotating} shows a comparison between the non-rotating case presented above and rotating cases with $\text{Ta} \in \{1,5\}(\times 10^{8})$. 
We see that the addition of rotation has no significant influence upon the evolution of $\left<u_x\right>$. 
In fact, we see that rotation tends to reduce the (already insignificant) integrated deviation from the target profile; rotation therefore seems to have a stabilising influence upon the system. 
As indicated by the lower part of Fig.~\ref{fig_hydrostability_rotating}, due to the presence of Coriolis effects, we see that weak mean flows emerge in the $y$-direction in the rotating cases. 
These could either be a transient response to the small-scale thermal perturbations that are added to the initial polytrope or the result of a very weak hydrodynamic instability. Either way, it should be stressed that these flows are rather low amplitude and are unlikely to have any significant impact upon the system over the timescales of interest. The inclusion of rotation has no significant (adverse) impact on the hydrodynamic stability of the system.

\section{Magnetic buoyancy with and without rotation}\label{section_main_results}

Having chosen a shear amplitude of $A=0.02$, we now impose a vertical background magnetic field (as described in \S~\ref{section_model}). 
As indicated in Table~\ref{table_dimensionless_parameters}, where all of the other parameters are specified, we still need to choose the values of $F$ (which determines the magnetic field strength) and the Taylor number (recall that we quote the mid-layer value, $\text{Ta}$).  
The choices for these parameters were guided by a preliminary low-resolution parameter survey, the details of which can be found in Appendix~\ref{appendixA}.

As well as using a somewhat weaker shear flow than \cite{silvers_double-diffusive_2009}, we also impose a weaker initial field, defined by $F = 2.5\times10^{-6}$ \citep[compared to $F\ge 1.25\times 10^{-5}$ in][]{silvers_double-diffusive_2009}.
This might seem to be a somewhat counter-intuitive decision, as this will increase the time taken for the formation of the horizontal magnetic layer, necessitating longer numerical calculations. 
However, this also delays the point at which the Lorentz force feedback starts to perturb the mean flow (the details of which will be described below).
Across the parameters surveyed, this reduced value of $F$ has a beneficial impact upon the magnetic buoyancy instability, because it allows steeper magnetic field gradients to develop. 
In what follows, we will present results for three different values of the Taylor number, corresponding to a non-rotating case ($\text{Ta}=0$) and two rotating cases, $\text{Ta} = 10^{8}$ and $\text{Ta} = 5 \times 10^{8}$.

\subsection{Non-rotating case}\label{section_non-rotating}
\begin{figure*}
    \includegraphics[width=2\columnwidth]{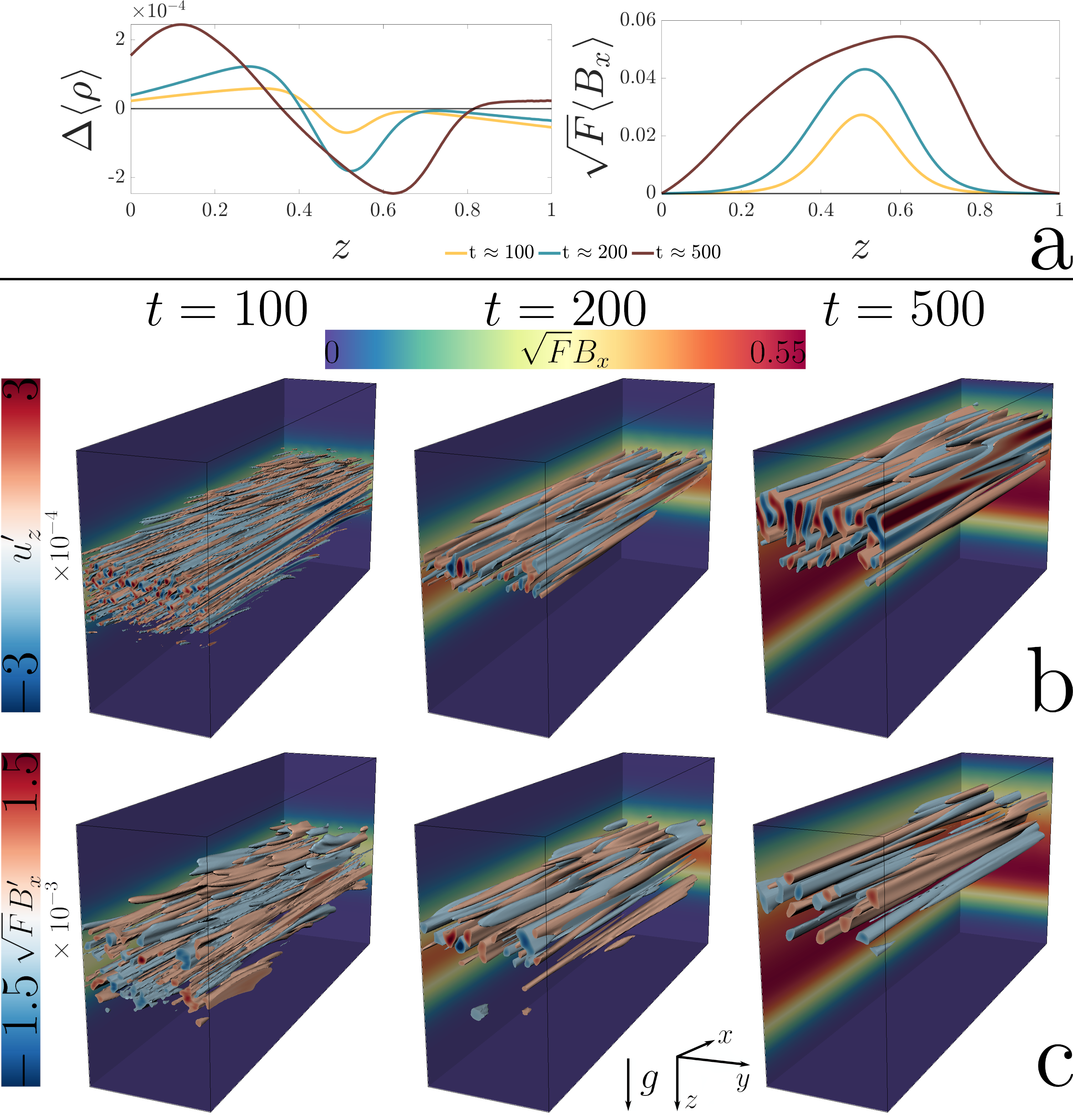}
	\caption{The evolution of the magnetic buoyancy instability for the non-rotating case. 
	Top section (a): Vertical profiles of the horizontally averaged density perturbation (left), defined as $\Delta\langle\rho\rangle = \langle\rho\rangle - \rho_\text{init}(z)$ (where $\rho_{\rm init}(z)$ is the initial, dimensionless density distribution of the polytropic layer), and vertical profile of the horizontally-averaged toroidal field $\sqrt{F}\langle B_x\rangle$ (right). 
	These are shown for three times in the evolution denoted by the legend. 
	The bottom section consists of 3D renderings at three times (see text labels) of pseudocoloured isovolumes (with associated legends to the left). The backdrop for these isovolumes shows 2D slices (with associated legend to the top)
	of the toroidal field $B_x$ at the associated times. 
	Middle row (b):  the vertical component of velocity $u_z$ with isovolumes defined by the regions $\lvert u_z \rvert \ge 10^{-4}$. 
	Bottom row (c): the toroidal field perturbation defined as $\sqrt{F}B_x^{\prime} = \sqrt{F}(B_x - \langle B_x \rangle)$ in isovolumes defined by the regions $\sqrt{F}\lvert B_x^{\prime} \rvert \ge 1.5\times 10^{-3}$.}
	\label{fig_flow_evolution}
\end{figure*}

Before considering the effects of rotation, we first give a detailed overview of the evolution in the non-rotating case,
the key features of which are illustrated in Fig.~\ref{fig_flow_evolution}.
In the early stages of the simulation, for $t \lesssim 100$,
the magnetic field is too weak to play a significant dynamical role.
During this stage there are only small-amplitude departures from the forced shear flow, which we attribute to acoustic and internal gravity waves that are excited by the initial temperature perturbation.
These oscillate with periods $\lesssim 10$ (dimensionless units), and dissipate on the thermal diffusion timescale, $\tau_\text{therm} = 100$.
At the same time, the imposed vertical magnetic field is continually stretched out by the shear flow, forming a magnetic layer around the mid-plane whose strength grows linearly in time. This stretching process is analogous to the $\omega$-effect at the base of the solar convection zone.

In Fig.~\ref{fig_flow_evolution} we also show the state of the system at $t=200$. 
By this time, the magnetic pressure from the mean toroidal field, $\langle B_x \rangle$, has become strong enough to noticeably change the overall mass distribution, moving fluid out of the shear region and into the layers above, as evident from the plots of $\Delta\langle\rho\rangle = \langle\rho\rangle - \rho_\text{init}(z)$, where $\rho_\text{init}$ is the initial density profile. 
This upward redistribution of mass provides the energy source for the magnetic buoyancy instability.
At the onset of magnetic buoyancy, we see 
tube-like structures in the isovolumes that are aligned with the shear direction, evident in both the magnetic field and the flow. 
We illustrate these in Fig.~\ref{fig_flow_evolution} through the vertical flow component and the toroidal field perturbation. 
It is clear that $B_x^{\prime}$, which is defined as  $B_x^{\prime} = B_x - \langle B_x \rangle$, and $u_z$ have structures that are localised around the upper regions of the induced magnetic layer, where the magnetic field gradient is conducive to magnetic buoyancy. 
As expected, we see that the magnetic buoyancy instability is characterised by a long length-scale parallel to the field and a short length-scale perpendicular to it.
In this case, this effect is perhaps exaggerated by the presence of the imposed shear, which will tend to promote ``near-interchange'' modes over those with a more undular profile \citep{tobias_influence_2004}.
These non-rotating results are qualitatively consistent with those of \citet{silvers_double-diffusive_2009}, albeit for a slightly different parameter regime.

\begin{figure*}
    \includegraphics[width=2\columnwidth]{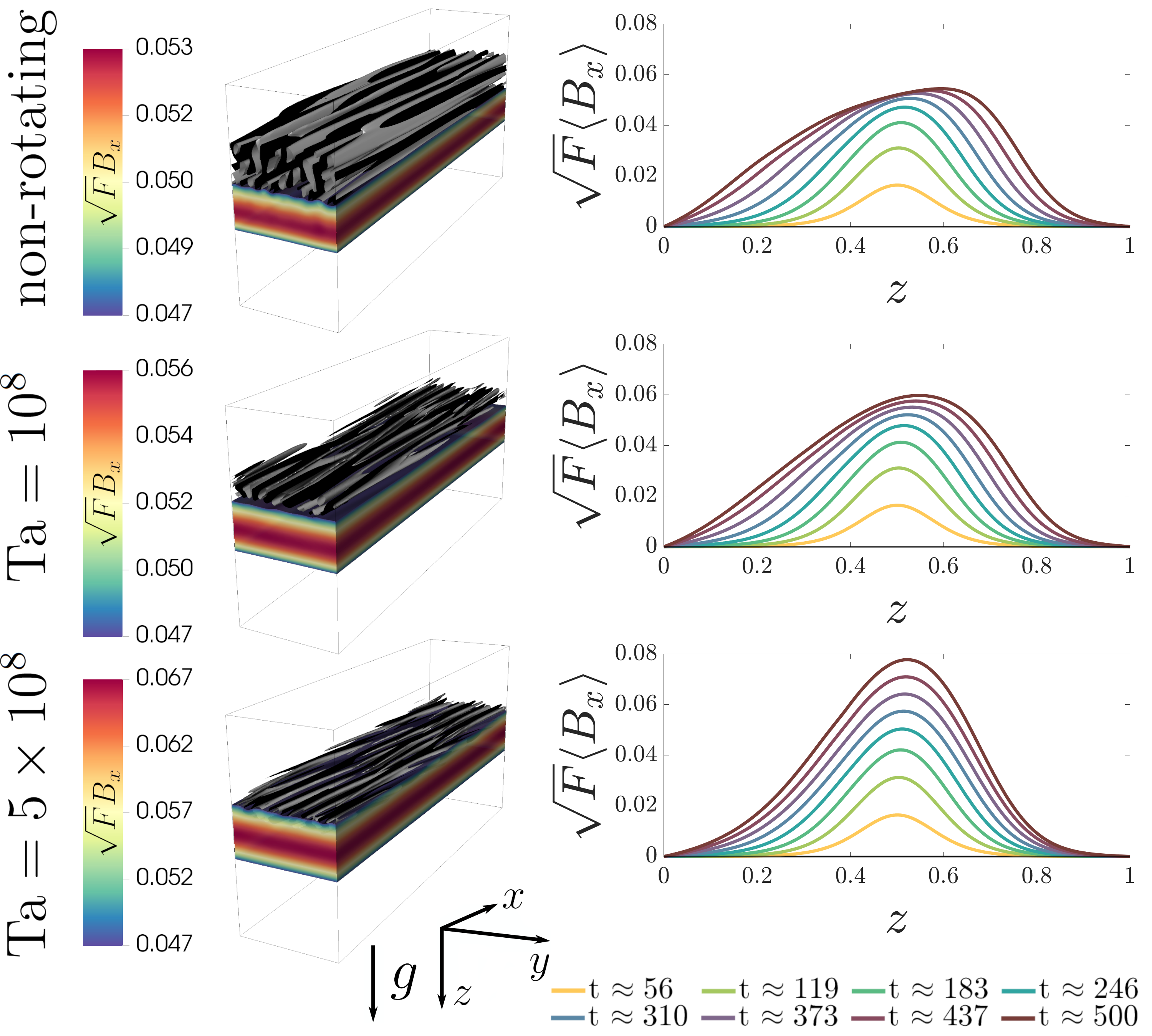}
	\caption{Magnetic buoyancy and mean magnetic field evolution for $\text{Ta}=0$, $\text{Ta}=10^8$ and $\text{Ta}=5\times 10^8$ (top to bottom). Left: volume renderings consisting of a pseudocoloured isovolume of $\sqrt{F}B_x$, defined by the region where $\lvert B_x \rvert > 30$, where the magnitude is denoted by the colour (see legend). Superimposed on this are isosurfaces of the vertical component of velocity $u_z$ defined at values $\pm 2.5\times10^{-4}$ (dark indicates positive and light indicates negative). Each snapshot is taken at $t\approx 400$ non-dimensional time units. Right: time snapshots of the horizontally averaged vertical profiles of the induced toroidal field $\sqrt{F}B_x$. Each case contains a number of snapshots taken at times denoted by the legend. }
	\label{fig_Bx}
\end{figure*}

As the simulation evolves, the buoyancy-driven motions become more pronounced and the non-trivial field/flow structures migrate towards the upper boundary of the computational domain.
This is clearly apparent at $t=500$ in Fig.~\ref{fig_flow_evolution}.
It is worth noting that there is no appreciable arching of the magnetic field structures as the instability evolves.
Eventually, the upper boundary condition starts to play a significant role in the dynamics of the system.
As we discuss in more detail below, the system also starts to exhibit significant (magnetically-driven) deviations away from the initial imposed shear flow. 
Motivated by these considerations, we therefore limit our integration times to $t\approx 500$ in all simulations.
As shown in Table~\ref{table_dimensionless_parameters}, the characteristic viscous and Ohmic timescales (based on the layer depth) are of the order of $10^5$ time units.
Given that these timescales are significantly longer than the evolution time for these simulations, it is clear that the observed migration of the field/flow structures is not a diffusively driven process. 

\subsection{The effects of rotation}\label{section_rotating}

Having set out the key features of this model in the non-rotating case, we now turn our attention to the effects of rotation. 
Throughout this subsection, quantitative comparisons are made between the non-rotating case ($\mathrm{Ta}=0$) and two rotating cases with $\mathrm{Ta}=10^8$ and $\mathrm{Ta}=5\times 10^8$. 

Fig.~\ref{fig_Bx} illustrates some of the effects of rotation upon this system, focusing upon the evolution of the vertical velocity, $u_z$, and the mean toroidal magnetic field, $\left<B_x\right>$.
Regardless of the rotation rate of the domain, there are many qualitative similarities between the rotating cases and the non-rotating case that is illustrated in Fig.~\ref{fig_flow_evolution}.
In particular the shear is always able to produce a magnetic layer that is susceptible to magnetic buoyancy. 
However, there are some significant differences that are attributable to the effects of rotation. 
In particular, as the rotation rate increases, we see that the development of the magnetic buoyancy instability is delayed.
The left-hand panel of Fig.~\ref{fig_Bx} shows the vertical velocity distribution at $t\approx 400$ for the three cases; whilst each exhibits clear indications of magnetic buoyancy, the associated flow structures (at this fixed instant in time) become progressively less well-developed as the rotation rate increases. 
One of the other notable features illustrated by this plot is that stronger toroidal fields
are 
generated at the mid-plane in the more rapidly-rotating cases. 
This is another indication of the delayed onset of the magnetic buoyancy instability, which would otherwise tend to disrupt this layer.
In cases where the magnetic buoyancy instability is more vigorous, we see a significant 
redistribution of the mean toroidal field, with a flattening of the peak around the mid-plane.

\begin{figure}
    \includegraphics[width=\columnwidth]{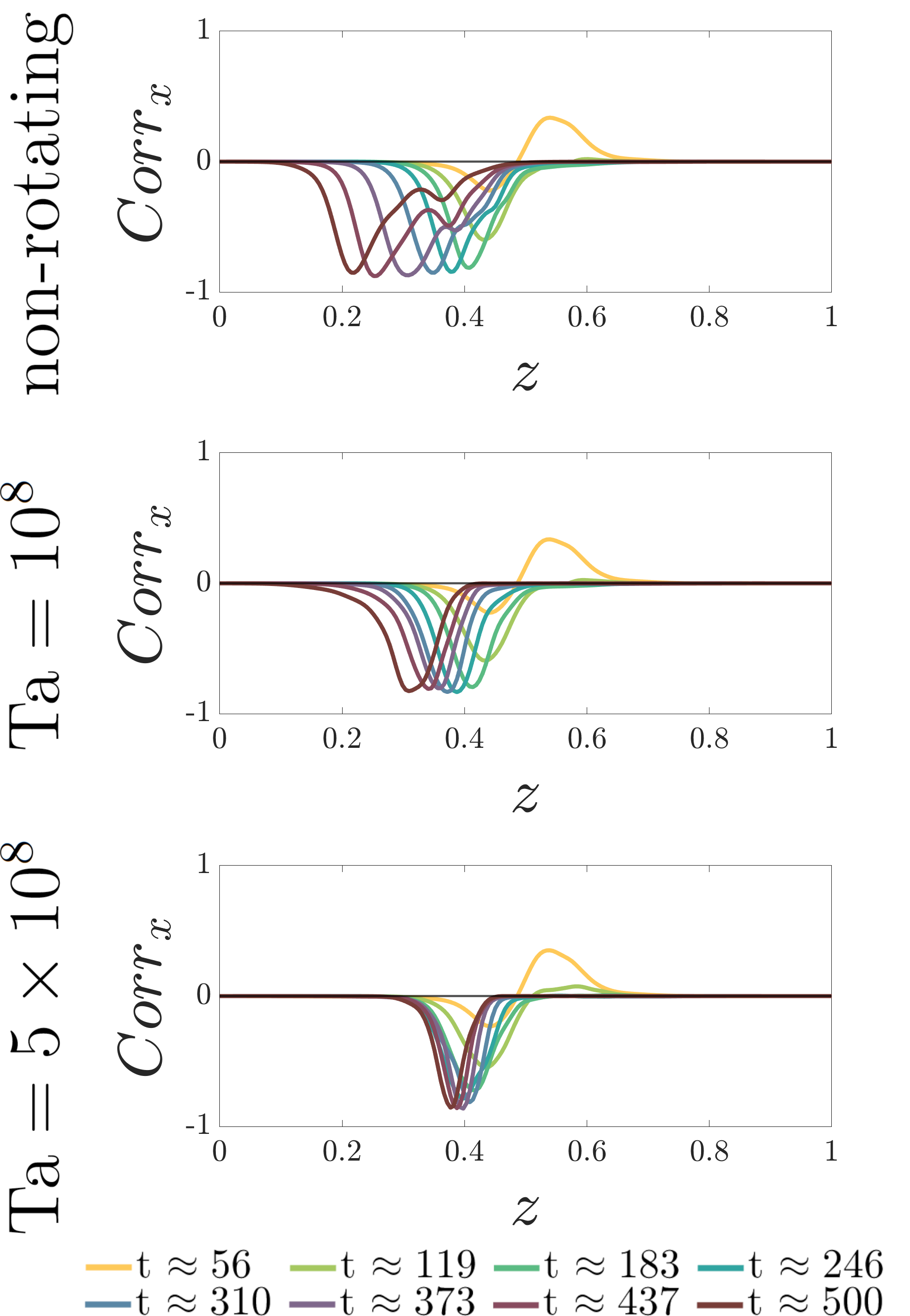}
	\caption{Vertical profiles of the correlation, $\text{Corr}_x(z)$, between $u_z$ and $B_x$, as defined by Eq.~(\ref{maths_correlation}) for $\text{Ta}=0$, $\text{Ta}=10^8$ and $\text{Ta}=5\times 10^8$ (top to bottom), where each plot contains a number of snapshots taken at times denoted by the legend.}
	\label{fig_corr}
\end{figure}

To investigate the rotational-dependence of the onset of the magnetic buoyancy instability in a more quantitative manner, we compute the Pearson correlation between the vertical flows and the induced horizontal magnetic field. 
Following \cite{silvers_double-diffusive_2009}, we define
\begin{align}\label{maths_correlation}
	\text{Corr}_x(z,t) = & \frac{\langle u_z B_x\rangle - \langle u_z \rangle \langle B_x \rangle}{\max_z\left[\sqrt{   \langle u_z^2\rangle - \langle u_z\rangle^2}    \sqrt{\langle B_x^2\rangle - \langle B_x\rangle^2}\right]} \,,
\end{align}
and then plot this quantity as a function of depth and time in Fig.~\ref{fig_corr} (for the non-rotating and the two rotating cases). 
We emphasise that this quantity does not have a straightforward physical interpretation (although, as we shall describe later, it is related to one of the terms that generate the $y$-component of the mean EMF). 
Nevertheless, it can be regarded as an indicator of magnetic buoyancy, because regions of stronger than average horizontal magnetic field are likely to be associated with buoyant upflows.
In the presence of magnetic buoyancy, we would therefore expect $\text{Corr}_x$
to be negative, and this is borne out by Fig.~\ref{fig_corr}.
In all cases shown, after an initial transient phase, we observe significant negative correlations in this quantity above the mid-plane of the domain.
The depth at which this correlation is maximal moves towards the surface as $t$ increases.
However, the rate at which this peak moves is strongly dependent upon the rotation rate.
In the non-rotating case, this peak has reached $z\approx 0.2$ by $t\approx 500$; at $\text{Ta}=5\times 10^8$, this peak is closer to $z\approx 0.4$ at the same time. 
This is generally consistent with the delayed development of magnetic buoyancy-induced motions in the rotating cases.

\begin{figure}
    \includegraphics[width=\columnwidth]{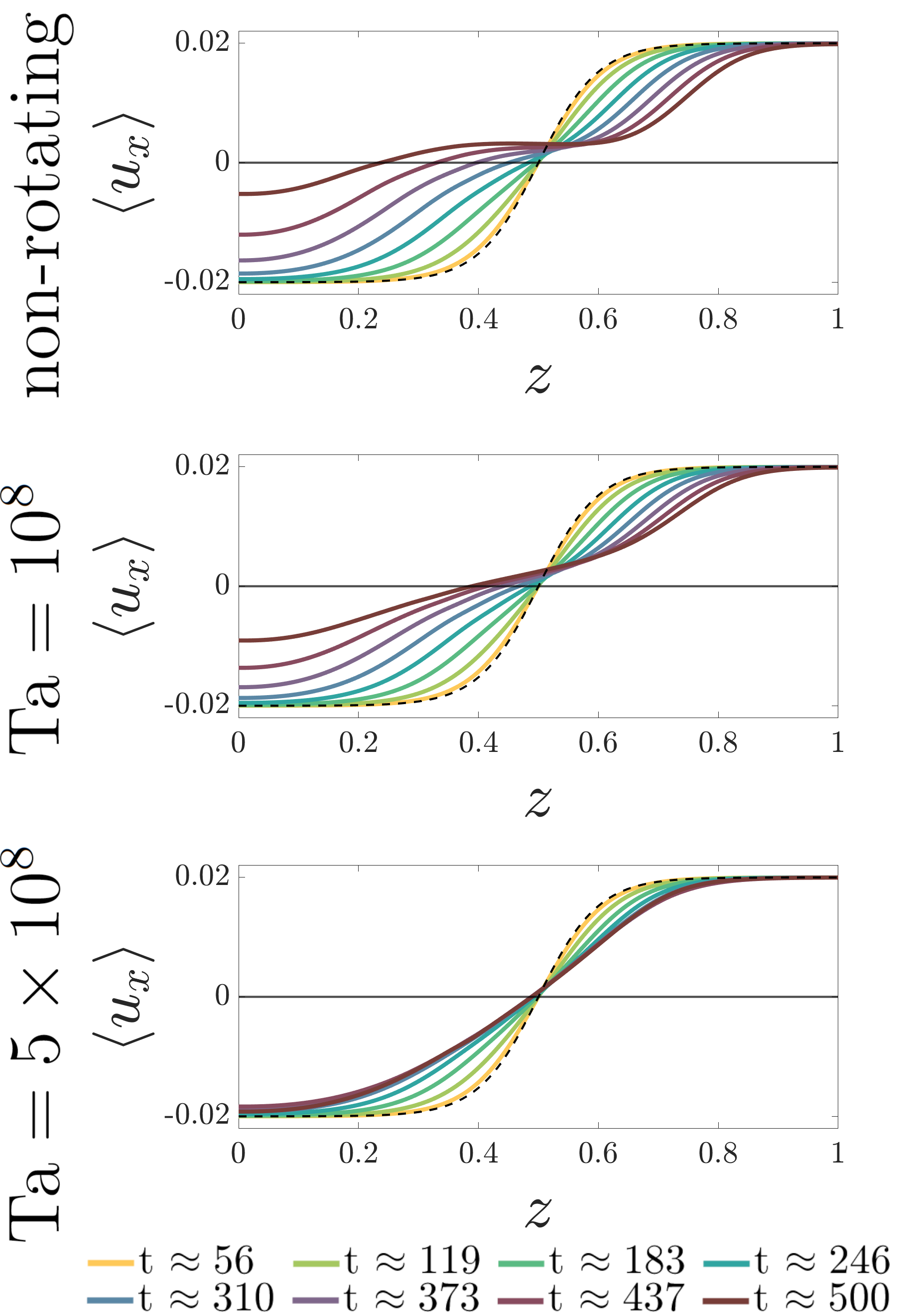}
	\caption{Vertical profiles of the horizontally-averaged $u_x$ component of velocity for $\text{Ta}=0$, $\text{Ta}=10^8$ and $\text{Ta}=5\times 10^8$ (top, middle, and bottom respectively), and all other parameters defined in Table~\ref{table_dimensionless_parameters}. 
	Each case has snapshots of the shear profile taken at various times denoted by the legend. 
	The dashed black line highlights the target shear profile.}
	\label{fig_ux_profiles}
\end{figure}

The effects of rotation are also evident in the evolution of the shear flow.
Fig.~\ref{fig_ux_profiles} shows the evolution of $\langle u_x\rangle$ for the three different cases. 
In all cases, we see a gradual deviation away from the initial shear.
This is due to the effects of the Lorentz force associated with the growing mean toroidal magnetic field, which will tend to resist the stretching due to the shear. 
Initially this Lorentz force feedback simply leads to a weakening and broadening of the shear profile, which is still largely concentrated about the mid-plane.
Once magnetic buoyancy develops, the shear profile flattens about the mid-plane and the resultant momentum redistribution leads to a significant perturbation to the flow structure, particularly at the upper boundary (where the density is relatively low). 
Longer integrations lead to further departures from the initial shear profile.
As the rotation rate increases, the deviation of the flow profile away from the initial state becomes more gradual. 
Indeed, in the most rapidly rotating case the shear flow still remains localised around the mid-plane even at $t\approx 500$. 

\begin{figure*}
    \includegraphics[width=2\columnwidth]{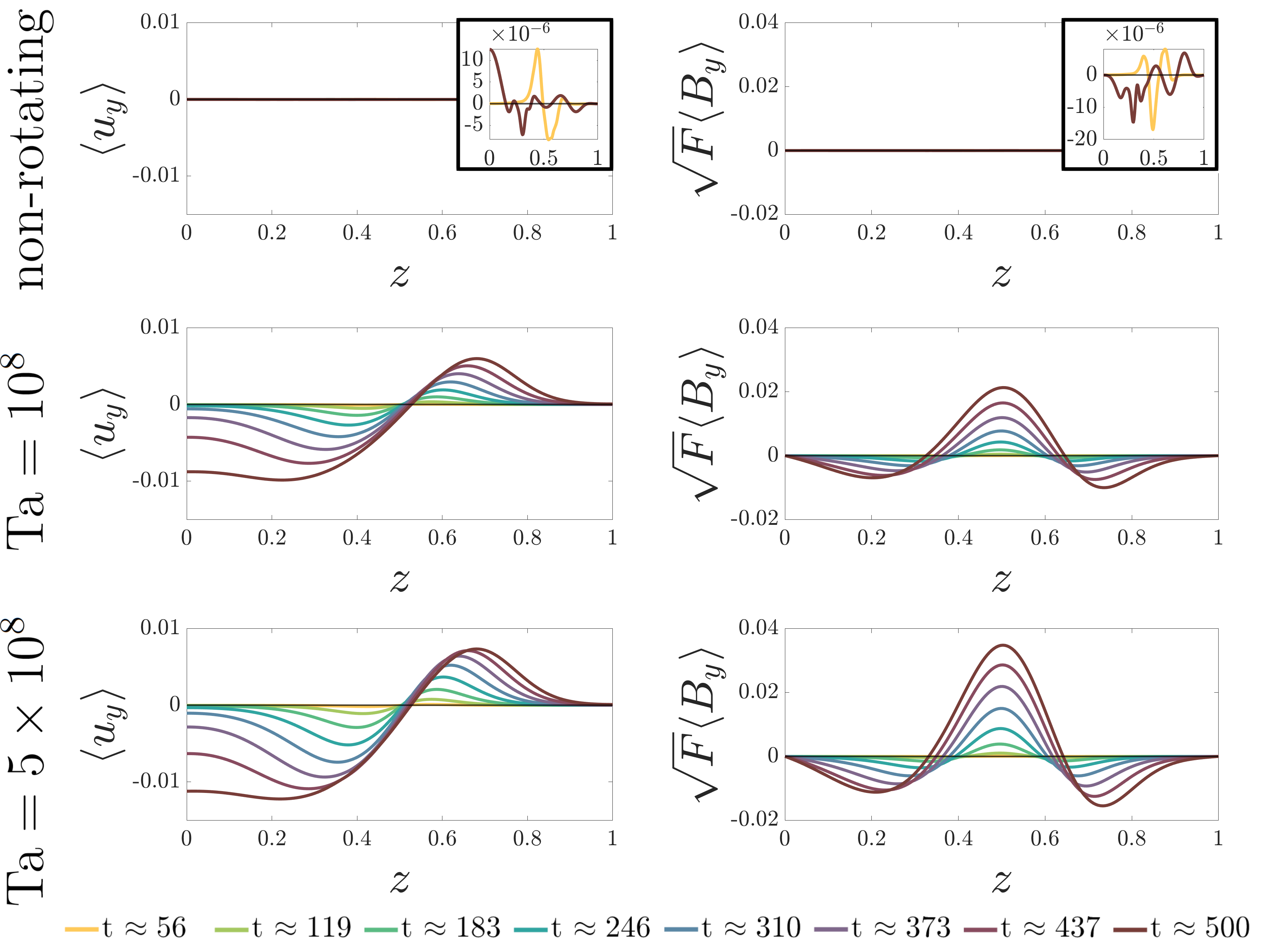}
	\caption{Time snapshots of the horizontally-averaged vertical profiles of the $u_y$ component of velocity (left) and the $B_y$ component of the magnetic field (right) for $\text{Ta}=0$, $\text{Ta}=10^8$ and $\text{Ta}=5\times 10^8$ (top to bottom). Each case contains a number of snapshots taken at times denoted by the legend.}
	\label{fig_uyBy}
\end{figure*}

Another important feature of the rotating cases is the emergence of mean flows in the $y$-direction (which are not a feature of the non-rotating case). 
This is illustrated in the left-hand panels of Fig.~\ref{fig_uyBy}, which plot the time and depth dependence of $\langle u_y\rangle$ for the three cases under consideration.
In the rotating cases, such flows
arise because the Lorentz force from the growing magnetic field disturbs the balance between the imposed forcing, $\mathbf{F}_s$,
and the Coriolis and viscous forces.
Because the density is lower in the upper part of the domain, the strongest mean flows in the $y$-direction tend to occur in this region, where (towards the end of the calculation) they reach values that are of a similar order of magnitude to the peak mean flow in the $x$-direction. 
We recall that our boundary conditions impose zero tangential stresses on the upper and lower boundaries, and so there is no flux of horizontal momentum through these boundaries.
Therefore, the mean flows arise from an internal redistribution of momentum within the domain,
and the total momentum is conserved.

\begin{figure}
    \includegraphics[width=\columnwidth]{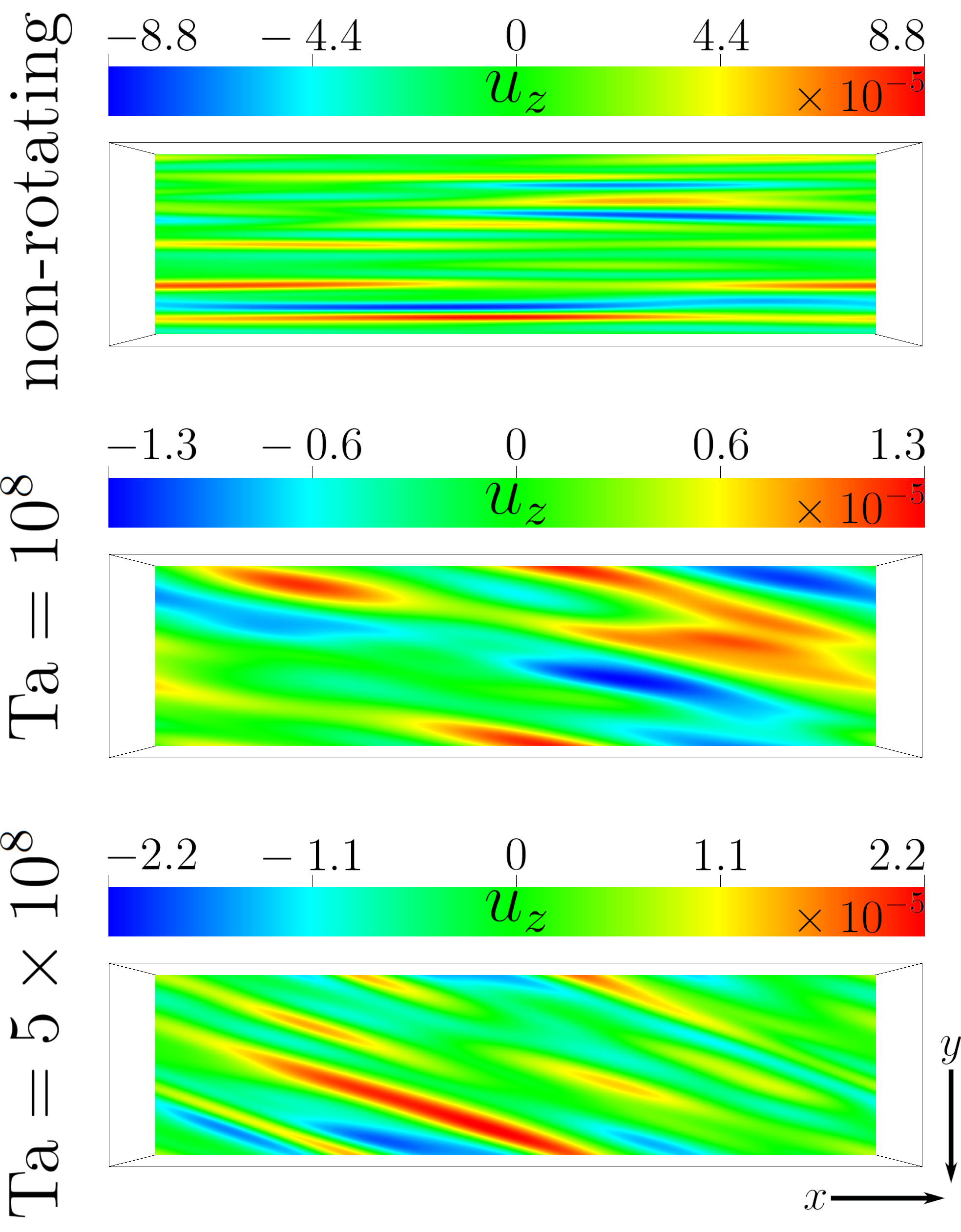}
	\caption{Plots of the $u_z$ component of velocity, with magnitude denoted by the colour bar, for $\text{Ta}=0$, $\text{Ta}=10^8$ and $\text{Ta}=5\times 10^8$ (top to bottom). Each plot is taken at the mid-plane, $z=0.5$, from a snapshot at $t\approx 400$ where the tilting effects, or lack thereof, of rotation can be most clearly seen.}
	\label{fig_planform}
\end{figure}

The right-hand panels of Fig.~\ref{fig_uyBy} show the time and depth dependence of $\langle B_y\rangle$. As expected, this quantity is practically zero in the non-rotating case. However, both rotating cases show the appearance of a systematic $\langle B_y\rangle$, which is an inevitable consequence of the behaviour of $\langle u_y\rangle$ that was discussed in the previous paragraph: the strong gradient in this quantity at the mid-plane stretches the vertical magnetic field into the $y$-direction. The peak value of $\langle B_y\rangle$ remains less than half of the peak value of $\langle B_x\rangle$, so the mean magnetic field is still predominantly in the $x$-direction. Nonetheless, the rotation of the mean magnetic field away from the $x$-axis affects the structure of the magnetic buoyancy instability. This is illustrated by Fig.~\ref{fig_planform}, which shows snapshots of the vertical flow, at the mid-plane of the domain; the tilting of these structures due to the effects of rotation is clearly apparent. 

\section{The mean EMF}\label{section_mean_EMF}

\begin{figure*}
    \includegraphics[width=2\columnwidth]{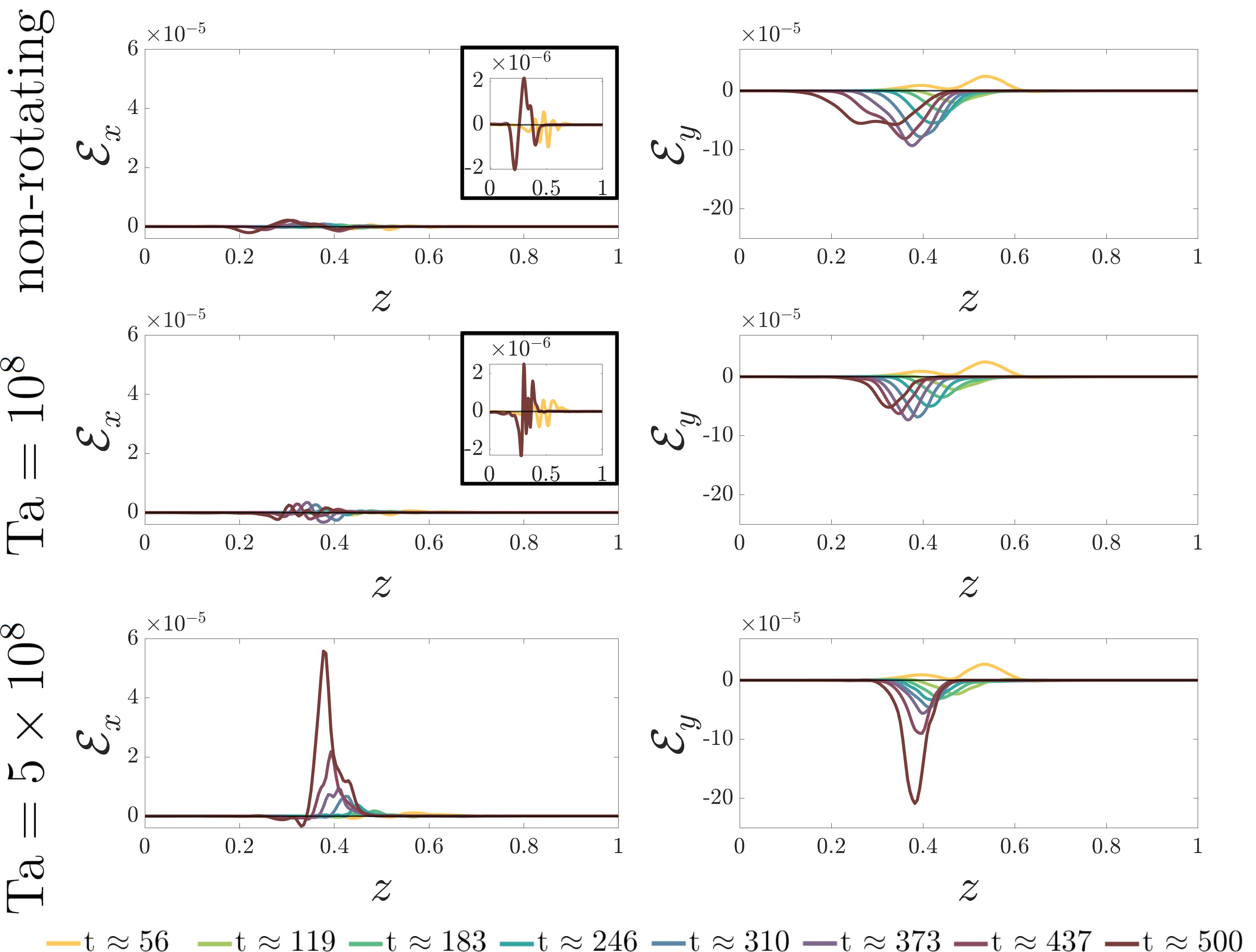}
	\caption{Plots of the mean EMF components $\mathcal{E}_x$ (left) and $\mathcal{E}_y$ (right), as a function of $z$, for $\text{Ta}=0$ (top row), $\text{Ta}=10^8$ (middle row) and $\text{Ta}=5\times 10^8$ (bottom row).
	The curves shown correspond to the times denoted by the legend. 
	Note that the axis ranges are held fixed as $\mathrm{Ta}$ is varied in order to facilitate quantitative comparisons between the three cases. (Insets are provided in two panels to illustrate the details of the low-amplitude variations.)}
	\label{fig_EMFs}
\end{figure*}

We now turn our attention to the most important result of this work, which is the analysis of the mean electromotive force (mean EMF) that is produced by the magnetic buoyancy instability in the presence of rotation.
According to the ideas originally put forward by \citet{parker_hydromagnetic_1955}, if the instability can produce a mean EMF with a significant component that is parallel to the mean magnetic field,
then 
this system might be able to drive an $\alpha \omega$-type dynamo. 
Before discussing the results from these simulations, we first review some of the key theoretical ideas. 

\subsection{Theoretical background}\label{subsection_EMF}

Following the standard procedures of mean-field theory, we can decompose the magnetic field and the velocity field into their mean and fluctuating parts.
To be specific, we can write $\mathbf{B} = \langle \mathbf{B} \rangle + \mathbf{B}^{\prime}$, where (as before) $\langle \mathbf{B} \rangle$
corresponds to the horizontally-averaged magnetic field, which is then a function of $z$ and $t$ only, whilst $\mathbf{B}^{\prime}$ represents the fluctuating component.
By construction, $\langle \mathbf{B}^{\prime} \rangle=\mathbf{0}$.
Similarly, we can express the velocity field in the form, $\mathbf{u} = \langle \mathbf{u} \rangle + \mathbf{u}^{\prime}$.
Applying the averaging operator to the induction equation (Eq.~\ref{maths_induction}), we see that

\begin{align}
	\dfrac{\partial\langle\mathbf{B}\rangle}{\partial t} =  
	\nabla \times \Big(\langle\mathbf{u} \rangle\times \langle\mathbf{B}\rangle
	+ \mathbf{\mathcal{E}}\Big) + \zeta_0 \kappa \frac{\partial ^2}{\partial z^2} \langle\mathbf{B}\rangle\,, \label{maths_meanfield_induction}
\end{align}
\noindent where we have defined the mean EMF
\begin{align}\label{maths_turb_emf}
	\mathbf{\mathcal{E}} = \langle\mathbf{u}^{\prime} \times \mathbf{B}^{\prime}\rangle \,.
\end{align}
The $\nabla \times \mathbf{\mathcal{E}}$ term in Eq.~(\ref{maths_meanfield_induction}) is the key ingredient of any mean-field dynamo model; under certain conditions it acts as a source term for the mean magnetic field. 
Due to the fact that $\mathbf{\mathcal{E}}$ is a function of $z$ and $t$ only, it is only the $x$ and $y$ components of the mean EMF that contribute to 
the evolution of the mean magnetic field. 
In what follows, we therefore restrict our analysis to these two components. 

In order to understand the potential influence of this mean EMF upon the dynamo, it is useful to review some of the key ideas from mean-field dynamo theory \citep[see, e.g.,][for more details]{moffatt_magnetic_1978}. 
For the moment, we make the simplifying assumption that there is an imposed (but potentially time-dependent) flow that evolves independently of the magnetic field, in order to consider the induction equation in isolation. 
It should be stressed that this assumption is not applicable for magnetic buoyancy, with the subsequent flow inextricably linked to the magnetic field; nevertheless, it is an assumption that allows insights to be gained into the properties of the mean EMF that might be conducive to dynamo action. 
We shall also assume that the mean quantities vary on a much longer length-scale than the fluctuations. 

Unless the imposed flow has the ability to generate a small scale dynamo (allowing the fluctuating magnetic field to grow in the absence of a mean field), there should be a linear relationship between the mean EMF and the mean magnetic field (and its derivatives). 
If the underlying flow were 
to be homogeneous and isotropic, then we could posit 
an expansion of the form 
\begin{align}\label{maths_turb_emf_expansion}
	\mathbf{\mathcal{E}} = \alpha\langle\mathbf{B}\rangle - \beta \nabla \times \langle\mathbf{B}\rangle + \ldots\,,
\end{align}
where $\alpha$ and $\beta$ are scalars (under more general conditions these would be tensorial quantities).
Under these simplifying assumptions, the $\alpha\langle\mathbf{B}\rangle$ term corresponds to the $\alpha$-effect and is non-zero only if the flow lacks reflectional symmetry (which is typically the case in the presence of rotation). 
In fact, under the first-order smoothing approximation, it is possible to derive an expression for $\alpha$ that is directly proportional to the kinetic helicity of the flow\footnote{We have not found an obvious relationship between the kinetic helicity of the turbulence and the mean EMF in our simulations and hence have not presented these results for brevity.}. 
We stress again that some of the simplifications made here are not really applicable in the context of magnetic buoyancy, and any interpretation of the mean EMF in terms of $\alpha$ and $\beta$ should be treated with a degree of caution \citep[][]{hughes_mean_2018,davies_mean_2011}.
In the following discussion, we shall therefore focus upon the properties of the mean EMF itself.
Nevertheless, these insights from mean-field theory suggest that the component of the mean EMF in the direction of the mean magnetic field has a significant bearing upon the potential for the system to act as a dynamo. 
Based upon these insights, we would also expect to see a strong dependence of this part of the mean EMF upon the rotation rate. 

Before moving on to analyse the mean EMF in these simulations, it is useful to relate this to previous work.
Two studies of particular relevance are those of \citet{davies_mean_2011} and \citet{chatterjee_alpha_2011}. 
In their linear analysis, \citet{davies_mean_2011} considered the mean EMF produced by magnetic buoyancy in an imposed, rotating, horizontal magnetic layer. 
Most of their analysis focused upon the case of an inclined rotation vector, but there are some general conclusions that are applicable to the vertical rotation case that is considered in the present paper. 
\citet{chatterjee_alpha_2011} carried out numerical simulations that are somewhat related to those of the present paper, although they considered the evolution of an imposed magnetic layer rather than a shear-generated one, carrying out an extended analysis of the latitudinal dependence of the system (largely at a fixed rotation rate).

Considering a unidirectional horizontal field, aligned with the $x$ direction, \citet{davies_mean_2011} found that the magnitude of the component of the mean EMF in the direction of the mean field ($\mathcal{E}_x$) does indeed increase with increasing rotation rate. 
Even in the absence of rotation, they also found a substantial component of the EMF in the horizontal direction that is perpendicular to the imposed magnetic field (i.e.\ $\mathcal{E}_y$), the amplitude of which decreased slightly as the rotation rate increased.
For the cases considered by \citet{davies_mean_2011}, they typically found that $\lvert\mathcal{E}_y\rvert \gg \lvert\mathcal{E}_x\rvert$.
Allowing for differences in the coordinate geometry of the two systems, the simulations of \citet{chatterjee_alpha_2011} exhibited similar behaviour, with the magnitude of the horizontal component of the mean EMF that is perpendicular to the imposed magnetic field exceeding that of the field-parallel component.
In both cases, this result can be traced back to the fact that the instability tends to be characterised by magnetohydrodynamical perturbations with small $y$-components. 

These imposed magnetic field studies provide us with some clear points of comparison, and we might expect to see some similar behaviour (in terms of the properties of the mean EMF) in the shear-generated case. However, given the greater complexity of our model, we should not be surprised if there are also some notable differences. 

\subsection{Numerical results}\label{subsection_EMF_numerics}

Fig.~\ref{fig_EMFs} shows the depth- and time-dependence of the mean EMFs for the three cases considered. 
We focus initially upon the $\mathcal{E}_x$ component. 
In the cases with $\text{Ta}=0$ and $\text{Ta}=10^8$, $\mathcal{E}_x$ takes both positive and negative values of similar magnitude, with rapid variations in $z$ and $t$, and displays no systematic trend.
In the most rapidly-rotating case ($\text{Ta}=5\times10^8$), by contrast,
$\mathcal{E}_x$ is positive for almost all values of $z$ and $t$,
with an amplitude that grows superlinearly in time.
Moreover, the location of the peak value of $\mathcal{E}_x$ migrates upward in time, roughly mirroring the migration of the buoyancy instability, as illustrated in Fig.~\ref{fig_corr}.
By $t\approx 500$, the peak value of $\mathcal{E}_x$ is an order of magnitude larger than the the peak value of $\lvert \mathcal{E}_x\rvert$ in either of the other two cases. 
Whilst this is clearly a more complicated model, this is qualitatively consistent with the results of \citet{davies_mean_2011}, who (as noted above) also found that this component of the mean EMF tended to increase with increasing rotation rate.

Turning our attention to $\mathcal{E}_y$, we find (after an initial transient phase) that in each case there is a significant component of the mean EMF in the negative $y$ direction,
whose magnitude is roughly an order of magnitude larger than that of $\mathcal{E}_x$
(this is qualitatively consistent with the results of \citealt{davies_mean_2011} and \citealt{chatterjee_alpha_2011}).
Interestingly, the $\text{Ta}=10^8$ case has slightly smaller values of $\mathcal{E}_y$ than the non-rotating case.
However, in the more rapidly-rotating case (with $\text{Ta}=5\times 10^8$), $\mathcal{E}_y$ is initially smaller,
but eventually grows to be significantly larger,
roughly in proportion with $\mathcal{E}_x$.
This non-monotonic dependence of the $\mathcal{E}_y$ on $\text{Ta}$ probably reflects the fact that the geometry of the mean magnetic field changes with rotation rate.
In the rotating cases, the mean magnetic field has significant components in both the $x$ and $y$ directions,
and so the distinction between $\mathcal{E}_x$ and $\mathcal{E}_y$ (in terms of their orientation with respect to the mean field) also becomes less meaningful.

\begin{figure}
    \includegraphics[width=\columnwidth]{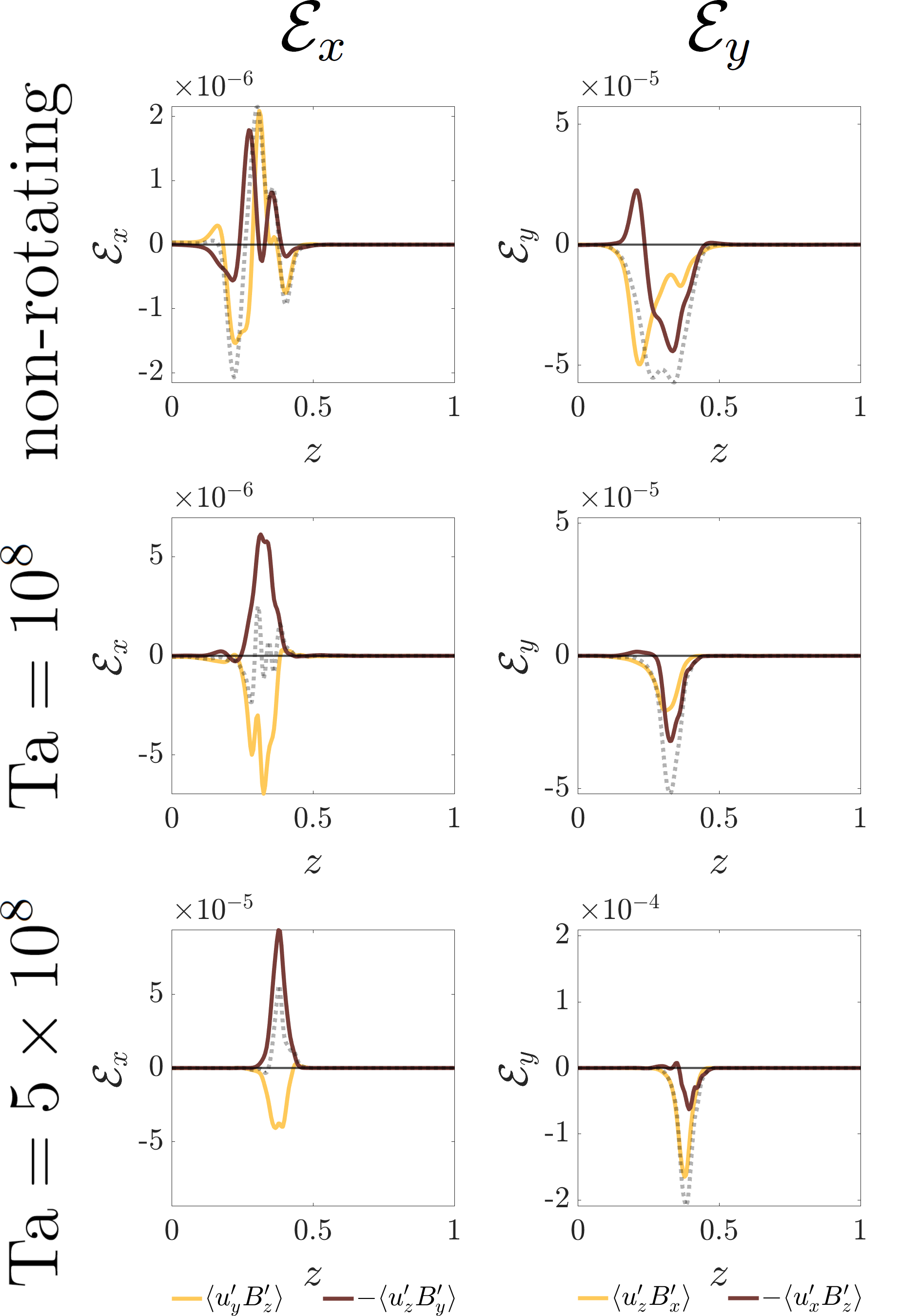}
	\caption{Comparisons between the individual terms, denoted by colour (see legend), that make up the EMF components $\mathcal{E}_x$ (left) and $\mathcal{E}_y$ (right). The comparisons are made for an individual time snapshot taken at $t\approx 500$. We make these comparisons for $\text{Ta}=0$ (top row), $\text{Ta}=10^8$ (middle row) and $\text{Ta}=5\times 10^8$ (bottom row). For reference we show the total $\mathcal{E}_x$ and $\mathcal{E}_y$ in each case as a dotted line.}
	\label{fig_EMF_components}
\end{figure}

In Fig.~\ref{fig_EMF_components} we carry out a more detailed analysis of the components of the mean EMF at $t\approx 500$, for each of the three cases. 
To be specific, we have considered separately the contributions of $\langle u_y^\prime B_z^\prime \rangle$ and $-\langle u_z^\prime B_y^\prime\rangle$ to $\mathcal{E}_x$ (and have carried out a similar decomposition for $\mathcal{E}_y$). 
In the $\text{Ta}=0$ case, there is little systematic behaviour evident in these plots; all quantities are of low amplitude, leading to a negligible overall $\mathcal{E}_x$. 
For $\text{Ta}=10^8$, both $\langle u_y^\prime B_z^\prime \rangle$ and $-\langle u_z^\prime B_y^\prime\rangle$ seem to have a well-defined depth dependence, with (approximately) equal and opposite magnitudes, leading to significant levels of cancellation when these quantities are added together. 
Whilst there is no significant net $\mathcal{E}_x$ in this case, the coherent depth dependence of these quantities does indicate that rotation is already playing an important dynamical role.
In the most rapidly rotating case, although the two contributing quantities still have opposite signs, they have very different magnitudes; a stronger contribution from $-\langle u_z^\prime B_y^\prime\rangle$ leads to a net $\mathcal{E}_x$ in this case.
Given the nature of the magnetic buoyancy instability, it is perhaps unsurprising that the contribution from $u_z^\prime$ should be playing a dominant role. 
Both the $\text{Ta}=0$ and $\text{Ta}=10^8$ cases have a well-defined $\mathcal{E}_y$ in which $\langle u_z^\prime B_x^\prime\rangle$ and $-\langle u_x^\prime B_z^\prime\rangle$ both play an important role.
(This means that the correlation defined in Eq.~(\ref{maths_correlation}) is generally not a reliable proxy for $\mathcal{E}_y$.) 
In the most rapidly rotating case, the $\langle u_z^\prime B_x^\prime\rangle$ term is dominant, leading to an increase in the magnitude of $\mathcal{E}_y$.


\begin{figure}
    \includegraphics[width=\columnwidth]{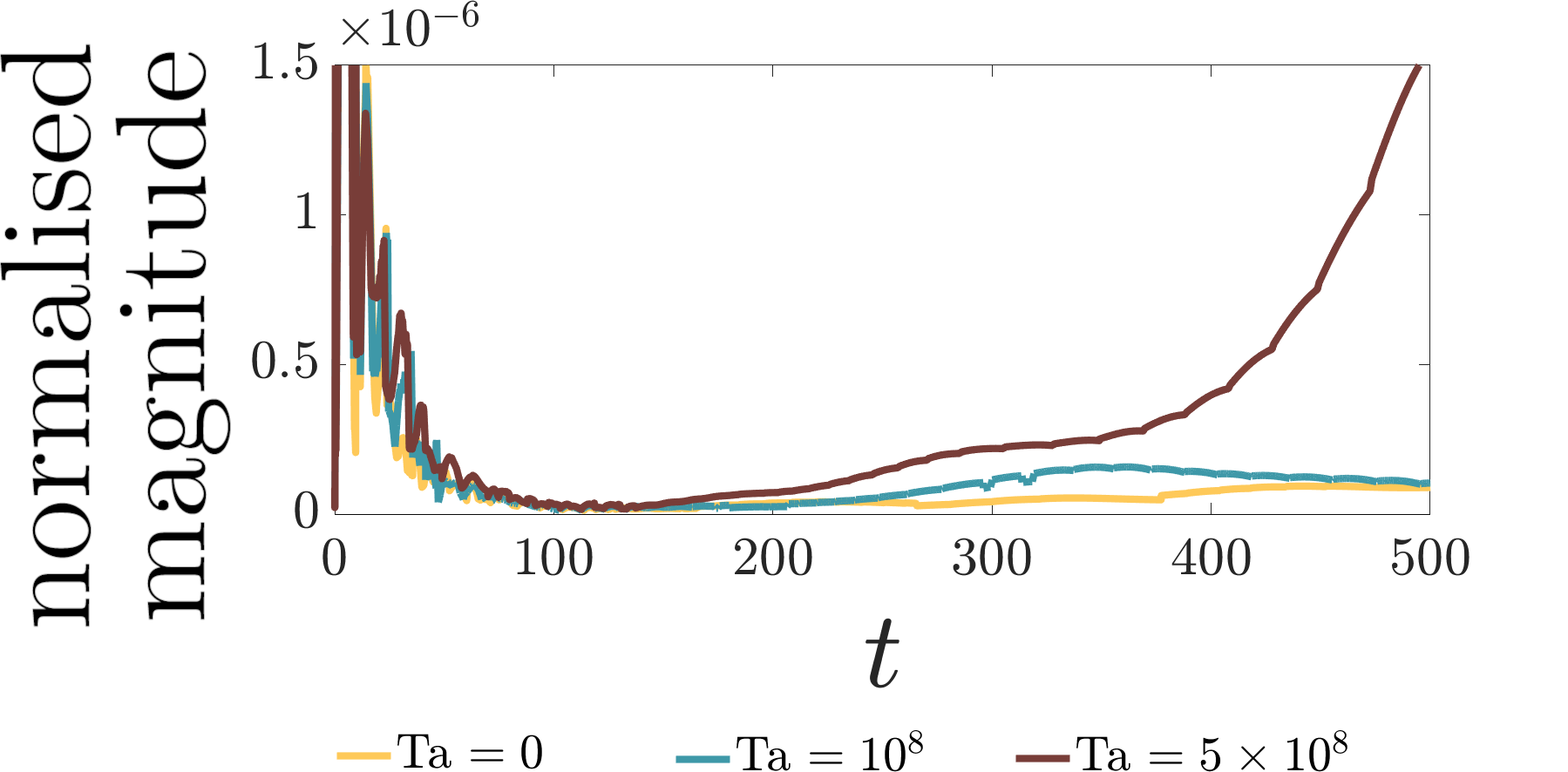}
	\caption{Time series of $|\mathcal{E}_x(z_m(t),t)|/\langle B_x\rangle(z_m(t),t)$ for the three rotation rates shown in the legend. }
	\label{fig_alpha}
\end{figure}

An obvious question to ask at this stage is whether or not these mean EMFs are large enough to have a significant regenerative effect on the magnetic field. 
Some insight can be gained by considering the relative magnitude of the mean EMF relative to the mean magnetic field, focusing particularly upon the $x$-component of this quantity (for the reasons presented above). 
However, this process is complicated by the fact that the relevant quantities are all depth-dependent.
To produce an indicative value of this normalised magnitude, we therefore define $z_m(t)$ to be the value of $z$ at which $\lvert\mathcal{E}_x\rvert$ takes its maximum value at that particular time. 
We then consider the quantity $\mathcal{E}_x(z_m(t),t)/\langle B_x\rangle(z_m(t),t)$. 
Whilst the definition of this normalised quantity is certainly motivated by the definition of $\alpha$ in Eq.~(\ref{maths_turb_emf_expansion}), we stress again that the validity of this expansion is questionable in this context, which explains why we choose to work with the mean EMF rather than describing this system in terms of $\alpha$.

In Fig.~\ref{fig_alpha}, we plot $|\mathcal{E}_x(z_m(t),t)|/\langle B_x\rangle(z_m(t),t)$, as a function of time, for the three cases ($\text{Ta}=0$, $\text{Ta}=10^8$ and $\text{Ta}=5 \times 10^8$). 
After initial transients have subsided, this normalised magnitude is comparable for each of the three cases, until the curves start to diverge at $t\approx 200$.
For both $\text{Ta}=0$ and $\text{Ta}=10^8$, there is no significant increase in this quantity after this point. 
In the more rapidly-rotating case, the normalised magnitude of the mean EMF increases rapidly, reaching a value of approximately $2 \times 10^{-6}$ at the end of the integration time. 
In dimensional terms, this normalised magnitude has the units of velocity, so it is natural to compare this quantity with the fluctuating component of the velocity field. 
As is evident from Fig.~\ref{fig_planform}, this peak value for the normalised mean EMF is approximately one order of magnitude smaller than the typical vertical velocity at the mid-plane of the domain. 
This is, therefore, still relatively small, but it should be noted that this quantity is still growing at the end of the simulation. 

In a classical $\alpha\omega$-dynamo model, it is perfectly possible for a dynamo to operate with a weak $\alpha$-effect provided that the shear is sufficiently strong. 
A full dynamo calculation would be needed in order to assess whether or not this EMF (in tandem with the shear) is capable of sustaining a Parker-like dynamo.
Moreover, such a calculation would require a much larger domain in the latitudinal direction
than the model presented here,
in order to allow for sufficient separation between the small-scale instability and the large-scale magnetic field.

\section{Discussion and conclusions}

Motivated by the original $\alpha\omega$-dynamo model that was first proposed by \cite{parker_hydromagnetic_1955}, the aim of this paper was to assess whether the interaction between magnetic buoyancy and rotation, in a shear-generated magnetic layer, can produce a regenerative effect for the mean magnetic field analogous to that of a convectively-driven $\alpha$-effect.
Our numerical model builds on that of \citet{vasil_magnetic_2008} and \citet{silvers_double-diffusive_2009}, who conducted simulations of a fully compressible magnetohydrodynamic fluid under the influence of a tachocline-like vertical shear flow.
An imposed vertical magnetic field is stretched by the shear to produce a predominantly horizontal field, which then becomes unstable to magnetic buoyancy. 
The novel aspect of the work presented here is the inclusion of rotation in this system, which breaks the reflection symmetry of the buoyancy-driven flows, potentially enabling this instability to supply the necessary mean electromotive force (EMF) required for a Parker-like dynamo to operate. 

To assess the influence of rotation, we carried out a detailed quantitative comparison of simulations with different Taylor numbers. 
Whilst rapid rotation does delay the development of the magnetic buoyancy instability, the imposed shear still leads to the formation of a magnetic layer that eventually becomes buoyantly unstable as the simulation progresses. 
If the rotation rate is high enough, the instability-induced perturbations (to the flow and magnetic fields) give rise to a systematic mean EMF, with a significant component, $\mathcal{E}_x$, in the direction of the mean magnetic field (see Fig.~\ref{fig_EMFs}). 
In agreement with previous analytical work \citep[e.g.][]{davies_mean_2011} we find that the magnitude of $\mathcal{E}_x$ increases with increasing rotation rate.
Like previous studies, we also find a significant $\mathcal{E}_y$ component of the mean EMF in all cases (regardless of the extent to which the layer is rotating).  

Following the ideas originally set out by \cite{parker_hydromagnetic_1955}, a systematic $\mathcal{E}_x$ (in combination with the imposed velocity shear) is likely to be conducive to dynamo action of $\alpha\omega$-type. 
However, the efficacy of any possible dynamo process will certainly depend on the magnitude of $\mathcal{E}_x$, which would typically be expressed in terms of an $\alpha$ coefficient in a standard mean-field dynamo model.
As noted above, $\mathcal{E}_x$ is certainly small (in a normalised sense) in all of these simulations, although it should be emphasised that it is growing throughout the final stages of the simulation in the most rapidly-rotating case (see Fig.~\ref{fig_alpha}).
It is therefore plausible that this regenerative term will eventually become large enough to influence significantly the evolution of the mean magnetic field (see Eq.~\ref{maths_meanfield_induction}).
Of course, some caution is needed when extrapolating results from any numerical simulations to conditions in the solar interior.
Furthermore, full dynamo calculations (without a uniform imposed magnetic field) are needed in order to confirm the viability of this dynamo mechanism.
Nevertheless, our initial results support the hypothesis that the magnetic buoyancy in the solar tachocline could be playing a crucial regenerative role in the solar dynamo.

An obvious limitation of our current simulations is that the mean flow eventually diverges from the tachocline-like shear profile that we seek to impose; this is the key determining factor in limiting the run time of these simulations to $t\approx500$. 
While this limited run-time does not detract from the results we have presented, it is worth noting that for a dynamo calculation it would be desirable to run these simulations for a much longer time, ideally a significant fraction of the Ohmic timescale (which for our chosen parameters is $\tau_\text{ohmic} \approx 2\times10^{5}$; see Table~\ref{table_dimensionless_parameters}).
The deviation from the imposed mean flow is an inevitable consequence of the Lorentz forces that result from the continual stretching of the imposed vertical field. 
This magnetic feedback perturbs the initial force balance, which smooths and then flattens the initial shear profile.
In the rotating cases this also drives mean flows in the $y$-direction.
Both of these changes to the mean flow are undesirable from the point of view of any dynamo calculation that seeks to mimic conditions in the solar tachocline.
One way to resolve this issue would be to employ a shearing box model, in order to remove the back-reaction of the Lorentz force onto the mean flow  \citep[e.g.][]{barker_magnetic_2012}. Alternatively an adaptive forcing could be incorporated, rather than the fixed forcing $\mathbf{F}_s$ used here, although this would inevitably introduce new dynamical effects into an already complex system.
In the absence of a complete understanding of the solar differential rotation,
any model of the tachocline's shear will necessarily be somewhat artificial.
Fortunately, in a full dynamo simulation, without an imposed vertical field,
the winding up of the horizontal field is likely to become a self-limiting process.
In that case, it may be possible to maintain a tachocline-like shear flow with the same kind of fixed forcing this is employed here.

Another potential issue with our existing model is that the buoyantly rising magnetic field eventually reaches the top of the domain.
At this point, the dynamics of the system start to become strongly dependent upon the choice of boundary conditions, which will never accurately replicate the conditions around the base of the solar convection zone (a fact that is true for any local model of the tachocline). 
A related point to note is that our magnetic boundary conditions allow magnetic fields to escape at the upper boundary, which might limit the efficiency of the dynamo if the field is able to escape faster than it can be regenerated in the tachocline.
Whilst we know that magnetic flux can escape from the tachocline, the rate at which this happens in any local model will certainly depend upon the precise boundary conditions adopted. 
The rise of the horizontal field towards the surface could be countered by magnetic pumping \citep{moffatt_magnetic_1978,moffatt_self-exciting_2019}, either via a parameterised pumping term in the induction equation \citep[e.g.][]{barker_magnetic_2012}, or by including a convective layer at the top of the domain \citep[e.g.][]{tobias_pumping_1998,tobias_transport_2001,brummell_penetration_2002,silvers_interactions_2009,weston_magnetic_2020}. 
We have started to carry out some preliminary studies on the viability of the convective analogue of magnetic pumping on the mean field, following a similar (parameterised) approach to that described by \cite{barker_magnetic_2012}, and plan to report on the results from these simulations in a future paper. If magnetic pumping, or the more technically challenging convective layer, can successfully pin down the toroidal field then it is possible this could increase the efficiency of any dynamo that this system might be able to excite. 

There are many other possible avenues for future work. 
All of the discussion in this paper has focused upon the case in which the rotation vector is aligned with gravity, which corresponds to the polar regions of the tachocline.
By considering an inclined rotation vector, it is possible to consider the latitudinal dependence of this system.
Given that active regions at the solar surface are confined to mid- to low-latitudes, this is a natural next step, not only for the imposed vertical magnetic field calculations that were considered in this paper, but also for the subsequent dynamo calculations.  
Another possible avenue of research is to explore possible sound-proof approximations to this system. In a parallel study, we are investigating the extent to which various sound-proof approximations adequately describe the linear onset of magnetic buoyancy in an imposed magnetic layer \citep{moss_validity_2022}.
Sound waves do not play a significant role in the dynamics of this system, so sound-proof approximations (which the relax the requirement of having to resolve the acoustic timescale) could allow significantly larger time-steps to be taken, thus increasing the possible simulation time. 
This could be highly beneficial for any future dynamo simulations.

\section*{Acknowledgements}
This work was supported by a Research Project Grant from the Leverhulme Trust (RPG-2020-109). This research made use of the Rocket High Performance Computing service at Newcastle University. 

\section*{Data availability}
The data underlying this article will be shared on reasonable request to the corresponding author.

\bibliographystyle{mnras}
\bibliography{mybib.bib}


\appendix
	
\section{Full list of simulations}
\label{appendixA}
For completeness we show the full list of simulations used in the analysis of this work. Table~\ref{table_simulation_list_production} lists the simulation which were used for the results presented in the main body of the paper, which had resolutions of $(N_x, N_y, N_z) = (192, 96, 192)$. These more computationally expensive production runs were guided by an extensive low resolution parameter sweep. 
In particular, a number of processes change the timescales of the relevant dynamics of the system, such as the reduction of shear amplitude and/or amplitude of $F$ acting to slow the rate of toroidal field production and hence increase computational cost. Similarly, it was unclear a priori how rapid the rotation needed to be in order to produce a significant mean EMF in the mean field direction.
The full list of low resolution simulations with $(N_x, N_y, N_z) = (128, 64, 128)$ can be found in Table~\ref{table_simulation_list}.

\begin{table}
\begin{center}
\begin{tabular}{ccc || ccc}
\begin{tabular}[c]{@{}c@{}}F\\ $(\times10^{-6})$\end{tabular}     & \begin{tabular}[c]{@{}c@{}}Taylor\\ $(\times10^{8})$\end{tabular} & A     & \begin{tabular}[c]{@{}c@{}}F\\ $(\times10^{-6})$\end{tabular}      & \begin{tabular}[c]{@{}c@{}}Taylor\\ $(\times10^{8})$\end{tabular} & A     \\ \hline \hline
0    & 1      & 0.02  & 2.5 & 0       & 0.02  \\
0    & 5      & 0.02  & 2.5 & 1       & 0.02  \\
0    & 0      & 0.02  & 2.5 & 5       & 0.02  \\
0    & 0      & 0.05  &          &         &      
\end{tabular}
\caption{A list of the field strength $F$, Taylor number, and shear amplitude $A$ for the set of simulations explored in this work. These cases each have numerical resolutions of $(N_x, N_y, N_z) = (192, 96, 192)$ and all other parameters are fixed by the values defined in Table~\ref{table_dimensionless_parameters}.}
\label{table_simulation_list_production}
\end{center}
\end{table}	

\begin{table}
\begin{center}
\begin{tabular}{ccc || ccc}
\begin{tabular}[c]{@{}c@{}}F\\ $(\times10^{-6})$\end{tabular}     & \begin{tabular}[c]{@{}c@{}}Taylor\\ $(\times10^{8})$\end{tabular} & A     & \begin{tabular}[c]{@{}c@{}}F\\ $(\times10^{-6})$\end{tabular}      & \begin{tabular}[c]{@{}c@{}}Taylor\\ $(\times10^{8})$\end{tabular} & A     \\ \hline \hline
0    & 0      & 0.02  & 3.75  & 0      & 0.015 \\
0    & 0      & 0.015 & 0.75  & 0      & 0.015 \\
0    & 0      & 0.05  & 1.875 & 0      & 0.015 \\
0    & 0      & 0.01  & 2.5   & 0      & 0.01  \\
0    & 0.1    & 0.02  & 2.5   & 0.1    & 0.02  \\
0    & 0.5    & 0.02  & 2.5   & 0.5    & 0.02  \\
0    & 1      & 0.02  & 5     & 1      & 0.02  \\
0    & 2      & 0.02  & 2.5   & 1      & 0.02  \\
0    & 3      & 0.02  & 1     & 1      & 0.02  \\
0    & 4      & 0.02  & 2.5   & 2      & 0.02  \\
0    & 5      & 0.02  & 2.5   & 3      & 0.02  \\
0    & 10     & 0.02  & 2.5   & 4      & 0.02  \\
12.5 & 0      & 0.05  & 5     & 5      & 0.02  \\
5    & 0      & 0.02  & 2.5   & 5      & 0.02  \\
1    & 0      & 0.02  & 1     & 5      & 0.02  \\
2.5  & 0      & 0.02  & 2.5   & 10     & 0.02 
\end{tabular}
\caption{A list of the field strength $F$, Taylor number, and shear amplitude $A$ for the set of low resolution simulations with $(N_x, N_y, N_z) = (128, 64, 128)$ that were used as an initial parameter sweep to guide this work. All other parameters are fixed by the values defined in Table~\ref{table_dimensionless_parameters}.}
\label{table_simulation_list}
\end{center}
\end{table}


\bsp	
\label{lastpage}
\end{document}